\documentclass[a4paper,fleqn]{cas-dc}

\usepackage[numbers,sort]{natbib}
\usepackage{verbatim}
\usepackage{algorithmicx,algorithm}
\usepackage{caption}
\usepackage{color}
\usepackage{ulem}
\usepackage{graphics,graphicx,url,color}
\usepackage{amsmath,amsfonts,amssymb,theorem,euscript,mathrsfs,bbm}
\newtheorem{theorem*}{Theorem}
\usepackage{epsf,epsfig,wrapfig}
\usepackage{array,enumitem,multicol}
\usepackage{footmisc}
\DeclareGraphicsExtensions{.ps}

\definecolor{navyblue}{rgb}{0.0, 0.0, 0.5}
\definecolor{lightblue}{rgb}{0.3, 0.3, 0.9}

\newcommand{\bracketprod}[2]{\left[{\kern-.75ex}\left[#1,#2\right]{\kern-.75ex}\right]_{}}

\DeclareMathOperator*{\argmin}{arg\,min}

\newcommand{\shah}{{\textstyle \amalg{\kern-4.pt\amalg}}}

\graphicspath{{./figs/},{./figs2/}}

\def\tsc#1{\csdef{#1}{\textsc{\lowercase{#1}}\xspace}}
\tsc{WGM}
\tsc{QE}
\tsc{EP}
\tsc{PMS}
\tsc{BEC}
\tsc{DE}

\begin{document}
\let\WriteBookmarks\relax
\def\floatpagepagefraction{1}
\def\textpagefraction{.001}
\shorttitle{A Global Constraint to Improve CT Reconstruction Under Non-Ideal Conditions}
\shortauthors{Ziyu Shu, Alireza Entezari}

\title [mode = title]{A Global Constraint to Improve CT Reconstruction Under Non-Ideal Conditions}                      



\nonumnote{This work was supported in part by the NSF [grant number CCF-2210866]}
\author{Ziyu Shu}[orcid=0000-0002-2530-3177]
\cormark[1]
\ead{ziyushu@ufl.edu}

\author{Alireza Entezari}
\address{CISE Department, University of Florida, Gainesville, FL 32611-6120, USA}



\cortext[cor1]{Corresponding author}

\begin{abstract}
\noindent {\bf Background and Objective:} The strong demand for medical imaging applications leads to the popularity of the CT reconstruction problem. Researchers proposed multiple constraints to tackle none ideal factors in CT reconstruction such as sparse-view, limited-angle, and low-dose conditions. Most of these constraints such as total variation and its variants are local constraints focusing on the relationship between a pixel and its neighbors. In this paper, we propose a new constraint utilizing the global prior of CT images to greatly reduce the streak artifacts and further improve the reconstruction accuracy.\\
\noindent {\bf Methods:} A CT image of the human body contains a limited number of different types of tissues, so pixels in CT images can be separated into several groups according to their corresponding types. In our work, we focus on the correct composition classification for individual pixels and utilize it as a global prior, which differs from priors utilized by most current constraints. We propose segmenting pixels based on their gray levels during the reconstruction process, and forcing pixels in the same group to have similar gray levels.\\	
\noindent {\bf Results:} Our experiments on the Shepp-Logan phantom and two real CT images from different benchmarks show that the proposed constraint can help the conventional local constraints further improve the reconstruction results under sparse-view, limited-angle, and low-dose conditions.\\
\noindent {\bf Conclusions:} Different from most current constraints focusing on the local prior, our proposed constraint only utilizes the global prior of CT images. In that case, our proposed constraint can collaborate with most local constraints and improve the reconstruction quality significantly. Furthermore, the proposed constraint also has the potential for further improvement, as the composition classification can be done with some more delicate methods, such as neural network related semantic segmentation algorithms. \\
\end{abstract}
\begin{keywords}
X-ray tomography, sparse-view, limited-angle, low-dose, compressed sensing, reconstruction algorithm. 
\end{keywords}
\maketitle

\section{Introduction}
\label{sec:intro}
CT reconstruction is one of the fundamental technologies in the medical imaging area. While the conventional filtered back projection algorithm and its variants are widely implemented in practice~\cite{pan2009commercial}, it cannot generate high-quality reconstruction results under non-ideal conditions. Model based iterative reconstruction (MBIR) algorithms are proposed to overcome this challenge. They formulate the image reconstruction as an inverse problem, and use a forward model to model the measuring process as well as non-ideal factors. However, under sparse measurement (sparse-view and limited-angle) and low-dose conditions, the measurements are either too few or too noisy for accurate reconstruction, making the inverse problem severely ill-posed. To tackle this problem, prior knowledge or assumptions were proposed as constraints to steer the algorithm during the iterative optimization process.

In recent years, the discovery from the field of compressed sensing showed that a signal could be recovered by incomplete measurement if it has a sparse presentation and the measurement is incoherent with the presentation bases. Furthermore, Candes et al.~\cite{candes2006near, candes2006robust} indicated that medical images could be approximately sparse by taking the magnitude of the gradient. This enables using total variation (TV) constraint and its variants~\cite{sidky2006accurate, jin2010anisotropic, wang2017reweighted} to improve the reconstruction accuracy.

Another successful implementation of the prior is discrete tomography, where the imaged object is assumed to consist of only a few different compositions, each represented by a constant gray level in the reconstruction. Prior knowledge of the gray values for each of the compositions can be used as a powerful constraint to generate a reconstruction that contains only these gray values~\cite{batenburg2011dart}. This prior can be applied to multiple tomography applications. Notable examples include the X-ray tomography of industrial objects, where we have the knowledge of their compositions beforehand. Also, a bone is assumed to have a single constant density if a micro-CT scanner is used.

In this paper, we slightly modify the prior of discrete tomography to further improve the reconstruction performance of general CT reconstruction problems. Instead of making discrete~\cite{batenburg2011dart} or partially discrete assumptions~\cite{roelandts2011pdart}, we propose that pixels in the reconstruction can be separated into several classes, each corresponding to a different body tissue represented by a specific range of gray levels. Although the statistics of gray levels (such as means or medians) cannot be used as a reconstruction result directly for the pixels in the corresponding group, they can be utilized as preliminary estimations during the iterative reconstruction process. Also, the proposed prior only focuses on the global grouping information, not the local relationship between a pixel and its neighbors. This indicates that the proposed constraint can better preserve sharp edges than local constraints. Furthermore, the proposed global constraint can work together with the conventional local constraints to further improve the reconstruction performance. The experiments on phantom and real CT images of different parts of the human body show the effectiveness of our proposed method under multiple circumstances.

The remainder of this paper is organized as follows: MBIR, streak artifacts, related priors, and the proposed method are introduced in Section \ref{sec:material}; the effectiveness of the proposed method is tested on phantom and real CT images with multiple widely used local constraints under different conditions in Section \ref{sec:exp}. The corresponding discussion is in Section \ref{sec:diss}, and the conclusion of this study is presented in Section \ref{sec:conclusion}.

\section{Materials and Methods}\label{sec:material}
\subsection{MBIR and Streak Artifacts}\label{sec:CT&MBIR}
Since its introduction in the 1970s, computed tomography (CT) has become an essential technology with a wide range of applications in the medical imaging area. Initially, due to the limitation of computation speed, researchers proposed the filtered back projection algorithm. It utilizes the equivalence between the Fourier transform of the reconstruction result and that of the measured sinogram signals to do CT reconstruction directly. However, the filtered back projection algorithm and its variants cannot always provide quality results as the aforementioned equivalence cannot take non-ideal factors into account (e.g. noise, projection geometry, and sparse measurement). MBIR algorithms overcome this challenge by using a forward matrix $\boldsymbol H$ to model the entire measuring system, converting the CT reconstruction problem to an inverse problem, which can be solved iteratively:
\begin{equation}
    \hat{\boldsymbol f} = \argmin_{\boldsymbol f}||\boldsymbol g - \boldsymbol H \boldsymbol f||^2_2,
\label{MBIR}
\end{equation}
where $\boldsymbol g$ indicates the measured sinogram signal, $\boldsymbol f$ indicates the attenuation map of the imaged object, $\boldsymbol{\hat{f}}$ indicates the reconstruction result, and the forward matrix $\boldsymbol H$~\cite{shu2020gram,shu2022exact} is used to model the entire measuring system, including but not limited to the projection geometry, discretization method, and sinogram sampling methods.

\begin{figure}[pos=htbp]
    \center
    \begin{minipage}[c]{0.45\linewidth}
        \centering
        \includegraphics[width=0.9\linewidth]{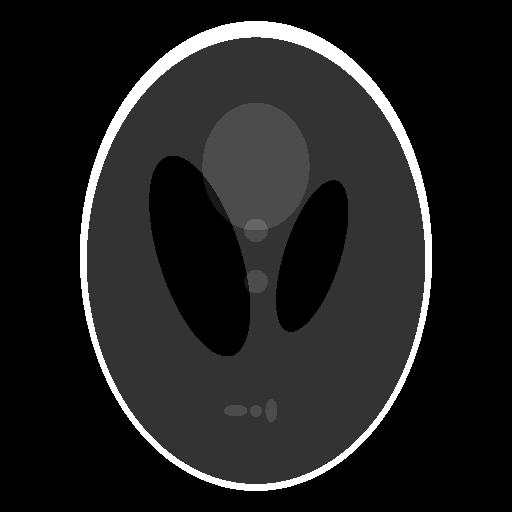}
    \end{minipage}
    \begin{minipage}[c]{0.45\linewidth}
        \centering
        \includegraphics[width=0.9\linewidth]{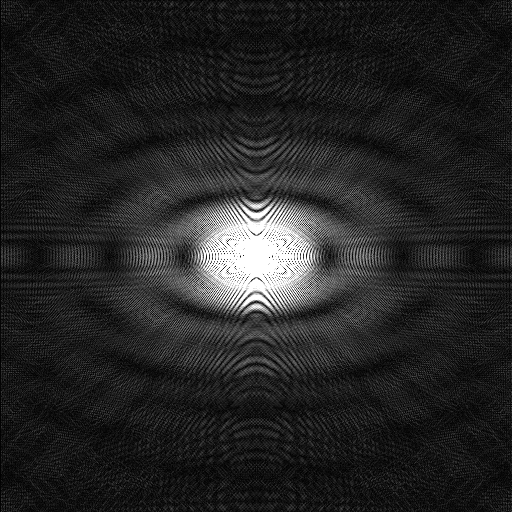}
    \end{minipage}
    \centerline{(a) Ground truth image}\medskip
    
    \begin{minipage}[c]{0.45\linewidth}
        \centering
        \includegraphics[width=0.9\linewidth]{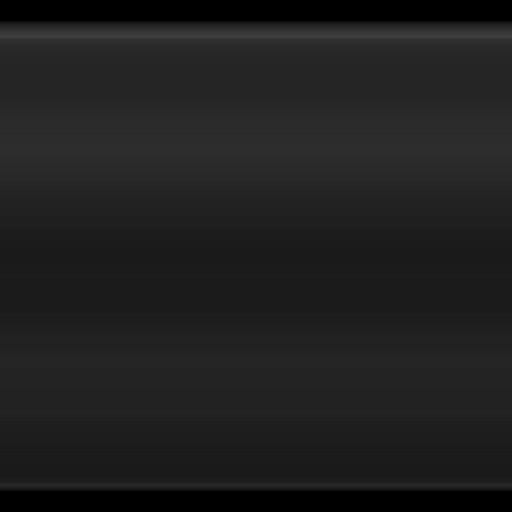}
    \end{minipage}
    \begin{minipage}[c]{0.45\linewidth}
        \centering
        \includegraphics[width=0.9\linewidth]{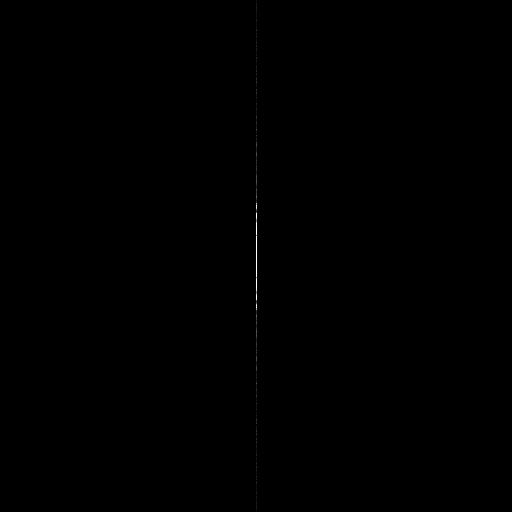}
    \end{minipage}
    \centerline{(b) Single-view back projection}\medskip

    \begin{minipage}[c]{0.45\linewidth}
        \centering
        \includegraphics[width=0.9\linewidth]{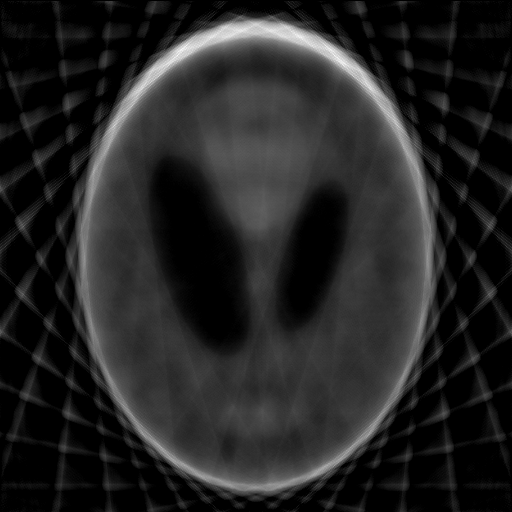}
    \end{minipage}
    \begin{minipage}[c]{0.45\linewidth}
        \centering
        \includegraphics[width=0.9\linewidth]{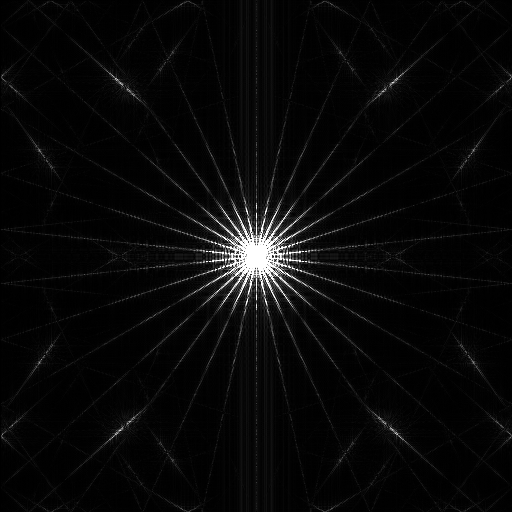}
    \end{minipage}
    \centerline{(c) Sparse-view reconstruction}\medskip
        
    \begin{minipage}[c]{0.45\linewidth}
        \centering
        \includegraphics[width=0.9\linewidth]{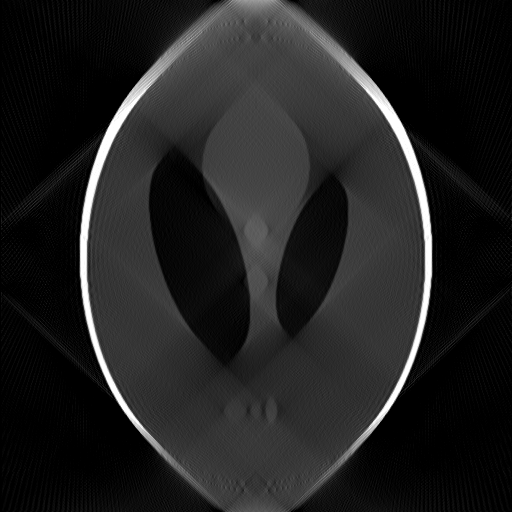}
    \end{minipage}
    \begin{minipage}[c]{0.45\linewidth}
        \centering
        \includegraphics[width=0.9\linewidth]{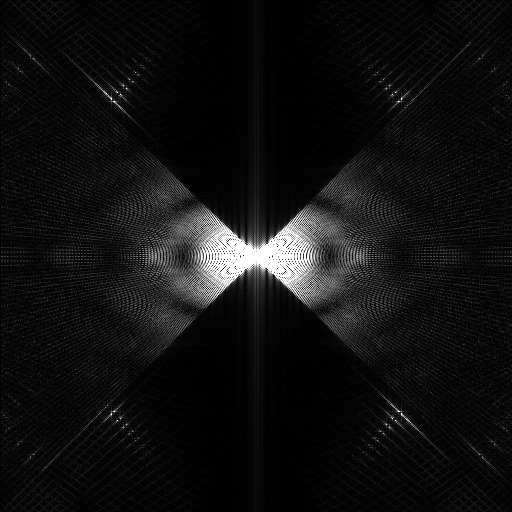}
    \end{minipage}
    \centerline{(d) Limited-angle reconstruction}\medskip   
    \caption{The illustration of streak artifacts in the image domain (first column) and frequency domain (second column). (a) the ground truth image. (b) single-view back projection. (c) the sparse-view (15 views distributed uniformly from $0^\circ$ to $180^\circ$) CT reconstruction. (d) the limited-angle (90 views distributed uniformly from $45^\circ$ to $135^\circ$) CT reconstruction. The apparent differences between the spectrum of (c) and (a), as well as (d) and (a) indicate the cause of streak artifacts.}
    \label{strakartifacts}
\end{figure}

It is worth mentioning that $\boldsymbol H^{\rm T}\boldsymbol H$ is a positive-definite matrix in CT reconstruction problems~\cite{shu2020gram,shu2022exact}, so that Equation \ref{MBIR} can be minimized iteratively, and its minimizer can be recursively determined:
\begin{equation}
\begin{aligned}
    &{\boldsymbol f}^k = {\boldsymbol f}^{k-1} + \alpha \boldsymbol H^{\rm T} \boldsymbol r,\\&\boldsymbol r = \boldsymbol g - \boldsymbol H{\boldsymbol f},
\end{aligned}
\label{MBIR2}
\end{equation}
where $\boldsymbol r$ is the inconsistency (residual) between the given measurement $\boldsymbol g$ and the modeled measure result of the current reconstruction result $\boldsymbol H \boldsymbol f$, $\alpha$ is the step size which is defined upon different algorithms. Equation \ref{MBIR2} indicates that the MBIR related algorithms back project the sinogram domain residual to the image domain in each of its iterations to minimize the data inconsistency. Since the back projection operation is the adjoint operator of the forward projection, its physical significance is to uniformly spread the sinogram value back along the rays measured in the forward projection. It is evident that data inconsistency can be minimized by this method. However, under the conditions that limit the number of views or the projection angular range, where multiple results can minimize the data inconsistency, the forward and back projection scenario cannot generate high-quality results as it will induce streak artifacts. A detailed representation is shown in Fig.\ref{strakartifacts}: the ground truth of the imaged object is shown in both the image and the frequency domain (Fig.\ref{strakartifacts}a); a single-view back projection is available in Fig.\ref{strakartifacts}b, showing the physical significance of back projection; sparse-view and limited-angle reconstruction results are shown in Fig.\ref{strakartifacts}c and Fig.\ref{strakartifacts}d respectively, where the streak artifacts are evident. It is worth mentioning that despite the existence of streak artifacts, the data inconsistency between the measurement and the forward projection of the reconstruction has already been minimized. The existence of artifacts is due to the incorrect prior implied in the back projection. It uniformly distributes the sinogram residual along the projection line, while the imaged object does not follow this prior. This can be better explained in the frequency domain. According to the central slice theorem, back projection can only update the frequency pixels on the frequency slices corresponding to the projection angles. As a clear example, Fig.\ref{strakartifacts}b shows that the single-view (projection angle $0^\circ$) back projection can only update one frequency slice and leave the rest frequency pixels to zero. As a result, reconstructions such as Fig.\ref{strakartifacts}c and Fig.\ref{strakartifacts}d are far from the ground truth (especially noticeable in the frequency domain), as we believe the ground truth must have a non-zero frequency spectrum in the unmeasured areas.

\subsection{Piece-Wise Constant Prior}\label{sec:TV}
As mentioned above, researchers proposed that medical images are approximately piece-wise constant. This enables the constraints to focus on the relationship between a pixel and its neighbors. The most widely used constraint is the total variation (TV)~\cite{sidky2006accurate}:
\begin{equation}
    \label{eq:TV}
    {\rm TV}(\boldsymbol f) = ||\nabla \boldsymbol f||_1 = \sum_{x,y}\sqrt{(\partial_x \boldsymbol f_{x,y})^2+(\partial_y \boldsymbol f_{x,y})^2},
\end{equation}
where $||\cdot||_1$ indicates the $l_1$ norm. Thus, TV is the summation of image gradient magnitude. It is a popular constraint and has been proven useful in improving the reconstruction accuracy under sparse measurement (sparse-view or limited-angle) and noisy conditions~\cite{sidky2006accurate, song2007sparseness, chen2008prior, sidky2008image}. 
Without loss of generality, a TV regularized reconstruction method can be expressed as:
\begin{equation}
\hat{\boldsymbol f} = \argmin_{\boldsymbol f}(||\boldsymbol g - \boldsymbol H \boldsymbol f||^2_2+\lambda ||\boldsymbol f||_{\rm TV}),
\end{equation}
where $\lambda$ indicates the weight of the constraint. When the system is complete or over-determined (the number of measurements is not less than that of unknowns), a unique result can be found for the term $||\boldsymbol g - \boldsymbol H \boldsymbol f||^2_2$ (data fidelity). In this case, the constraint only focuses on the trade-off between data fidelity and image regularity, and is mainly used for denoising. When the system is under-determined (e.g. sparse-view and limited-angle conditions), where multiple results are available for minimizing the data fidelity term, the constraint takes an extra role of choosing the most reasonable results from all candidates. In this case, the constraint is used to rectify the incorrect prior implied in the back projection scenario. 

The TV constraint can be further optimized for better performance under different conditions. For example, in the limited-angle CT reconstruction problem, where a certain range of viewing angles is missing, edges and details tangent to the projection directions are more likely to be recovered while blurs and artifacts only occur in some specific directions~\cite{quinto1988tomographic}. Furthermore, the contributions from blurred directions will weaken the effect of the TV constraint, since it is an isotropic summation of gradient magnitude. To overcome this challenge, Chen et al. and Jin et al.~\cite{chen2013limited, jin2010anisotropic} propose using anisotropic total variation (ATV) to eliminate the contributions from the blurred directions. One of its simple modifications can be written as:
\begin{equation}
    \label{eq:ATV}
    {\rm ATV}(\boldsymbol f) = ||\nabla_{a,b} \boldsymbol f||_1 = \sum_{x,y}\sqrt{a(\partial_x \boldsymbol f_{x,y})^2+b(\partial_y \boldsymbol f_{x,y})^2},
\end{equation}
where $a$ and $b$ are weights for vertical and horizontal partial derivatives. For example, in the case shown in Fig.\ref{strakartifacts}d, where the horizontal projections (projections from $0^\circ$ to $45^\circ$ and $135^\circ$ to $180^\circ$) are missing, a better result can be obtained by letting $a = 1$ and $b = 0.001$. This constraint returns to the standard TV when $a$ and $b$ are equal to $1$.

Another challenge is the use of the $l_1$ norm, which makes the TV constraint unable to remove streak artifacts efficiently while preserving sharp edges. An example is shown in Fig.\ref{strakartifacts}d. It is evident that the gradients around the clear edges are very high. On the other hand, the gradients around the artifacts are much lower, as the related pixel values decrease gradually. The best way to solve this problem is to use the $l_0$ norm, as it counts the number but not the magnitude summation of non-zero gradients. However, there is no applicable linear optimization algorithm due to its non-convex property. To overcome this challenge, an iterative reweighted technique has been proposed \cite{chartrand2007exact, candes2008enhancing, dong2013compressive,sidky2014constrained}. The researchers point out that the key difference between the $l_1$ and $l_0$ norm is the dependence on magnitude: larger coefficients are penalized more heavily in the $l_1$ norm than smaller coefficients, unlike the more uniform penalization of the $l_0$ norm. To address this imbalance, a weighted formulation of the $l_1$ norm is proposed. Its underlying principle is to obtain $l_0$ norm as a limit of $l_p$ norm when $p$ is close to $0$, and convert it into a series of reweighted TV minimizations, where each iteration solves a convex problem while the overall algorithm stays nonconvex. Incorporating ATV, Wang et al.~\cite{wang2017reweighted} propose the reweighted anisotropic TV (RwATV) constraint to further improve the limited-angle CT reconstruction, which can be expressed as:
\begin{equation}
    \label{eq:RwATV}
    {\rm RwATV}(\boldsymbol f) = ||\boldsymbol R \nabla_{a,b} \boldsymbol f||_1 = \sum_{x,y}r_{x,y}|\nabla_{a,b} f_{x,y}|,
\end{equation}
where $\boldsymbol R$ is the weight matrix, its element $r_{x,y}$ can be updated by using the following equation in each iteration:
\begin{equation}
    \label{eq:RwATV_r}
    r^{(l)}_{x,y} = \frac{1}{|\nabla_{a,b}f^{(l-1)}_{x,y}|+\epsilon},
\end{equation}
where $\epsilon$ is set to $10^{-8}$ to avoid a zero denominator. A trivial explanation about RwATV is to penalize the current gradients by the magnitudes of the gradients in the previous iteration. 

Besides the TV constraint and its variants, researchers also propose reconstructing the image by computing the maximum \textit{a posteriori} (MAP). By using the Markov random field as the prior model, the MAP method can be expressed as:
\begin{equation}
    \label{eq:qggmrf}
    {\rm min}_{\boldsymbol f}\sum_{\boldsymbol{s,r}\in \mathcal C} b_{\boldsymbol{s,r}}\rho (\boldsymbol{f_s-f_r}),\quad {\rm s.t.}\ \boldsymbol g = \boldsymbol H \boldsymbol f, \quad \boldsymbol f_{x,y} \geq 0,
\end{equation} 
where $\mathcal C$ is the set of all pairwise cliques, $b_{\boldsymbol{s,r}}$ are directional weighting coefficients, which are chosen as the inverse of the distance between the center pixel $\boldsymbol s$ and the elements in $\mathcal C$. $\rho(\cdot)$ indicates a potential function. Kisner et al.~\cite{kisner2012model} proposed using q-generalized Gaussian Markov random field (qGGMRF)~\cite{thibault2007three}, which has the form:
\begin{equation}
    \label{eq:qggmrf_p}
    \rho(\Delta) = \frac{{|\Delta|^p}}{1+|\Delta/c|^{p-q}},
\end{equation}
where $p$ and $q$ are used to determine the powers near, and distant from the origin respectively. The given function is close to $|\Delta|^p$ when $\Delta$ is small, and close to $|\Delta|^q$ when $\Delta$ is large. The parameter $c$ controls the transition point between these two cases. Especially, it contains special cases for some values of  $p$ and $q$:
\begin{itemize}
    \item Gaussian prior: $p=q=2$
    \item Approximate Huber prior: $p=2, q=1$
    \item Generalized Gaussian MRF: $1<q=p\le2$
    \item q-generalized Gaussian MRF: $1\le q<p\le2$
\end{itemize}

It is evident that the proposed potential function has the ability to smooth images while preserving edges if appropriate parameters are set.

\subsection{Discrete Tomography}\label{sec:DT}
In addition to the piece-wise constant prior, some more powerful priors can also be utilized under special conditions. A good example is the discrete tomography. In discrete tomography, the imaged objects are assumed to consist of a limited number of materials corresponding to different gray levels in the reconstruction. With such a powerful prior, discrete algebraic reconstruction technique (DART)~\cite{batenburg2011dart} is proposed, its basic steps are shown in Algorithm \ref{DART}. The method requires an initial guess for reasonable segmentation (line $2$). In each of its iterations (line $2$ to line $6$), pixels in the current result are separated into two groups (free pixels and fixed pixels). The free pixels will be updated by using conventional IR methods. It contains all boundary pixels between two different segments and some randomly selected non-boundary pixels. The random selection mechanism aims for dealing with noisy projection data and gray level errors. The fixed pixels are the rest of the pixels, indicating the interior part of the object. They can be updated directly based on the prior of the imaged object's composition. Their gray levels can be either known~\cite{batenburg2011dart} or calculated from the statistics of pixels' gray levels in the same group~\cite{van2012automatic}. As a result, high-quality reconstruction results can be obtained even if the system is highly under-determined.

The DART can be further modified to solve some non-discrete tomography problems. Partially DART (PDART)~\cite{roelandts2011pdart} assumes that only the densest material in the reconstruction is discrete. Van et al.~\cite{van2011optimal} propose accurately segmenting the dense homogeneous objects (metal) in the reconstruction and assigning gray levels to the corresponding pixels directly. In that case, both methods can keep the gray levels of the dense homogeneous objects fixed during the reconstruction process and thus reduce the noise and artifacts. However, parts in common imaged objects may not be completely homogeneous. Also, slight gray level inconsistencies, like those caused by inflammation, may be overlooked. Furthermore, in limited-angle CT reconstruction, accurate segmentation is unobtainable due to the strong artifacts along the edges of dense homogeneous areas.

Another fascinating development is the TVR-DART~\cite{zhuge2015tvr} algorithm, where the TV constraint is introduced for further improvement. This implies the feasibility of the collaboration between piece-wise constant prior and the priors related to discrete tomography. The reason is that the two priors are independent of each other: the former is a local prior focusing on the relationship between a pixel and its neighbors, and the latter is a global prior focusing on the correct segmentation.

\begin{algorithm}[htb]
    \caption{DART algorithm}
    \label{DART}
    \hspace*{0.02in} {\bf Parameters:} 
    initial guess $\boldsymbol c$
    \begin{algorithmic}[1]
        \State {\bf repeat:}
        \State \quad Compute the segmented image $\boldsymbol s$ from $\boldsymbol c$
        \State \quad Find the boundary pixels (set $B$) from $\boldsymbol s$
        \State \quad Define the free pixels (set $U$) as the union of $B$ and some randomly selected pixels
        \State \quad Define the rest of pixels as fixed pixels (set $F$)
        \State \quad Update $\boldsymbol c$: For pixels in set $F$, assign gray values according to their segmentation results; for pixels in set $U$, update their gray values using conventional IR methods.
        \State {\bf until} \{stopping criteria\}
        \State {\bf return} $\boldsymbol c$
    \end{algorithmic}
\end{algorithm}

\subsection{Nerual Network and Deep Image Prior}\label{sec:NN}
As a newly invented technique with great potential, the neural network can also be used in the CT reconstruction areas. It learns extra information (prior) from the training data to improve the reconstruction quality. Neural network related methods can be used directly, resulting in an end-to-end reconstruction framework such as~\cite{zhu2018image}, or indirectly, working as plugins for conventional reconstruction methods to improve their performance~\cite{shen2019low,shah2018solving}. However, an unavoidable challenge is the lack of a lower bound guarantee as these pre-trained models are optimized on the training dataset but not inference data. One potential method which can partially solve the problem is the inverse GAN technique~\cite{anirudh2019improving}, where the latent vector $\boldsymbol z$ of a pre-trained generator $G(\boldsymbol z)$ is optimized by solving the problem:
\begin{equation}
\begin{aligned}
    &\hat{\boldsymbol z} = \argmin_{\boldsymbol z}||\boldsymbol g - \boldsymbol H G(\boldsymbol z)||^2_2,
    \\
    &\hat{\boldsymbol f} = G(\hat{\boldsymbol z}).
\end{aligned}
\label{inverseGAN}
\end{equation}
Although a high-quality pre-trained model is still required, the inverse GAN technique guarantees that at least a local minimum of the objective function can be found in the space spanned by the generator $G$.

Recently, a new property of convolutional neural networks called deep image prior (DIP)~\cite{ulyanov2018deep} has been discovered. Researchers point out that the structure of the convolutional neural network itself is a powerful prior for generating natural images other than random noises. This implies that an untrained, randomly initialized generator $G$ can also be used in Equation \ref{inverseGAN} (in that case, both the input latent vector $\boldsymbol z$ and the weights of the neural network need to be optimized) to minimize the data inconsistency. 

One can analyze this technique from another aspect: conventional IR methods use linear optimization algorithms to minimize data inconsistency; DIP related methods~\cite{shu2022sparse} optimize a non-linear, randomly initialized convolutional neural network to minimize data inconsistency. The great ability of neural networks guarantees that the inconsistency can always be minimized, and the minimization itself indicates that the performance of the DIP methods is at least equivalent to that of conventional IR methods in terms of data inconsistency. The key difference here is the way of minimization. Conventional IR methods suffer from streak artifacts due to the use of back projection. DIP related methods will not generate streak artifacts as they use the backpropagation algorithm for minimization. Detailed introductions about the reconstruction process of DIP related methods are available from~\cite{shu2022sparse}.

Comparing with conventional reconstruction algorithms, these DIP related methods can generate much better results when the system is under-determined, but cannot easily generate detailed images. In other words, these methods cannot effectively improve their reconstruction accuracy as the number of measurements increases. Recently, this deficiency has been solved in the method called RBP-DIP~\cite{shu2022rbp} by using a U-net with a special residual back projection connection. In that case, the only challenge faced by the DIP related methods is the time complexity, as the reconstruction process is similar to training a model from scratch.

\subsection{Proposed Method}
\label{sec:METHODOLOGY}
\begin{figure}[pos=htbp]
    \center
    \begin{minipage}[c]{0.45\linewidth}
        \centering
        \includegraphics[width=0.9\linewidth]{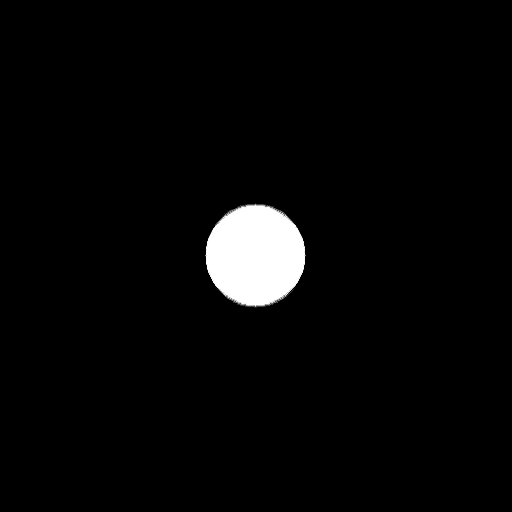}
        \centerline{(a) Ground truth}\medskip
    \end{minipage}
    \begin{minipage}[c]{0.45\linewidth}
        \centering
        \includegraphics[width=0.9\linewidth]{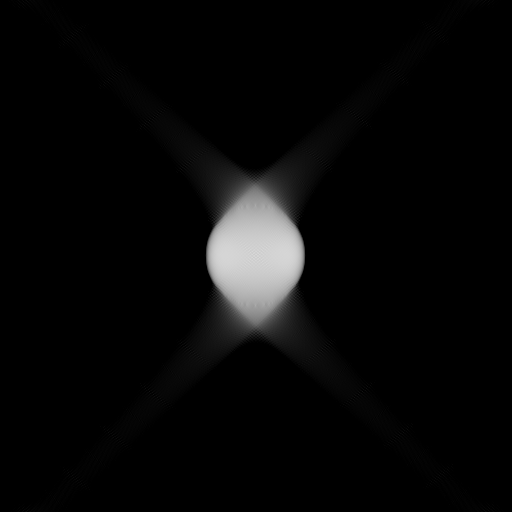}
        \centerline{(b) 50th iteration}\medskip
    \end{minipage}

    \begin{minipage}[c]{0.45\linewidth}
        \centering
        \includegraphics[width=0.9\linewidth]{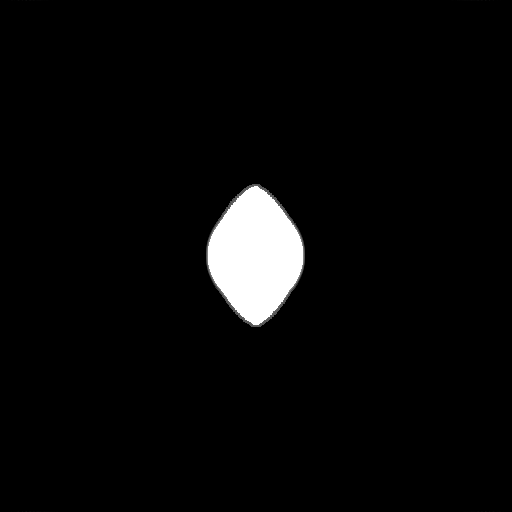}
        \centerline{(c) Segmentation of (b)}\medskip
    \end{minipage}
    \begin{minipage}[c]{0.45\linewidth}
        \centering
        \includegraphics[width=0.9\linewidth]{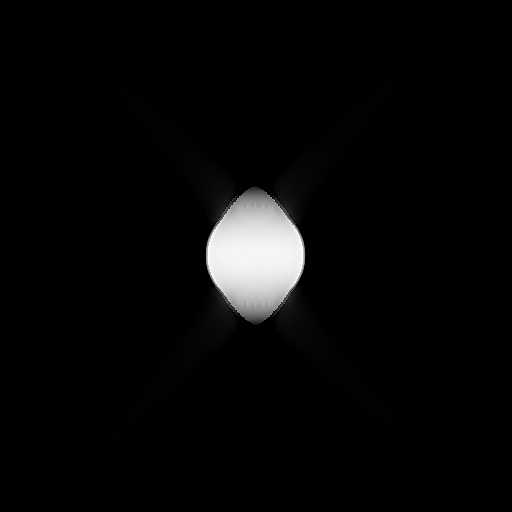}
        \centerline{(d) 100th iteration}\medskip
    \end{minipage}

    \begin{minipage}[c]{0.45\linewidth}
        \centering
        \includegraphics[width=0.9\linewidth]{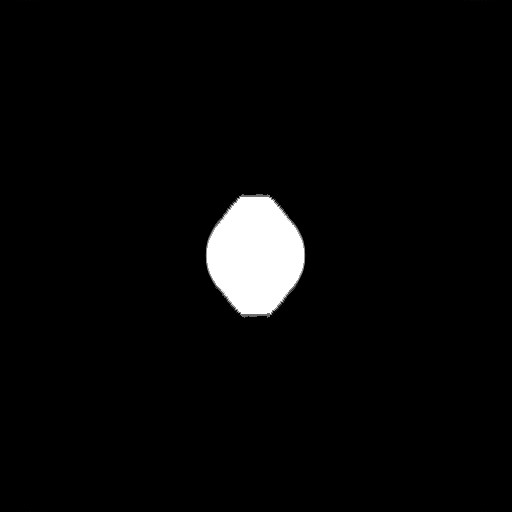}
        \centerline{(e) Segmentation of (d)}\medskip
    \end{minipage}
    \begin{minipage}[c]{0.45\linewidth}
        \centering
        \includegraphics[width=0.9\linewidth]{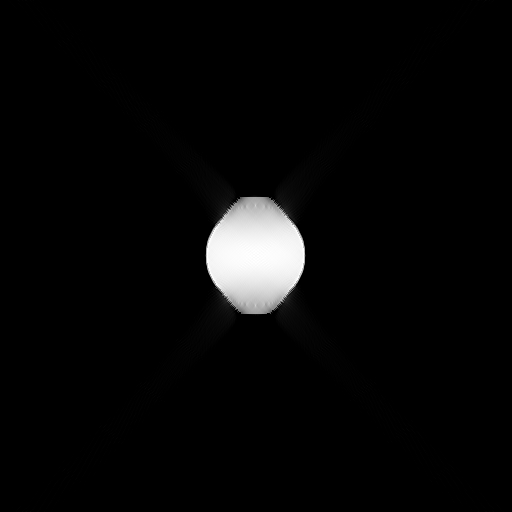}
        \centerline{(f) 500th iteration}\medskip
    \end{minipage}
        
    \begin{minipage}[c]{0.45\linewidth}
        \centering
        \includegraphics[width=0.9\linewidth]{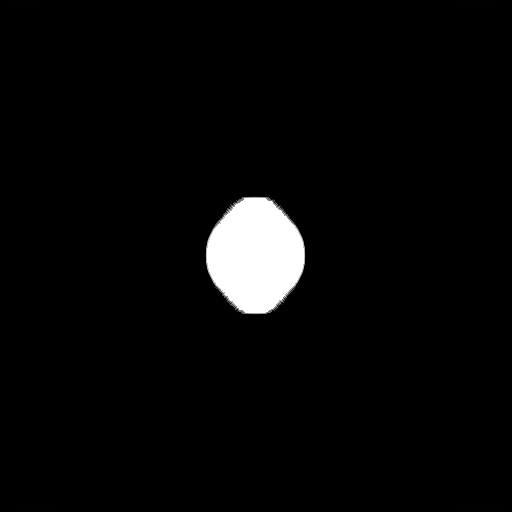}
        \centerline{(g) Segmentation of (f)}\medskip
    \end{minipage}
    \begin{minipage}[c]{0.45\linewidth}
        \centering
        \includegraphics[width=0.9\linewidth]{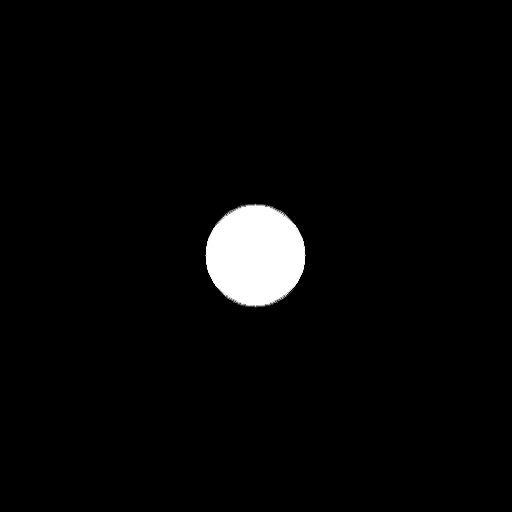}
        \centerline{(h) 1000th iteration}\medskip
    \end{minipage} 
    \caption{The illustration of artifact removal. (c), (e), and (g) are the segmentation of (b), (d), and (f) respectively. It is evident that the artifacts are being removed gradually.}
    \label{example}
\end{figure}

As shown in Fig.\ref{resultsp} and Fig.\ref{resultla}, there are mainly two types of errors in sparse-measurement CT reconstruction. The error along the edges is caused by drastic intensity change; the streak artifacts together with the missing boundaries are caused by the missing projection. The first columns of Fig.\ref{resultsp} and Fig.\ref{resultla} show the ground truth images used in this paper. It is evident that not only phantoms but also CT images of human beings consist of a few different compositions, each corresponding to a specific range of gray levels in the reconstruction, and this observation can be utilized as a prior. In that case, we propose approximately clustering the pixels into several groups, and enforcing the pixels in the same group to have a similar gray level during the reconstruction process. A brief explanation is shown in Fig.\ref{example}, where a circular phantom is imaged under the limited-angle condition (angular range $45^\circ$ to $135^\circ$). Due to the missing projections, artifacts generated by the back projection cannot be fully removed even when the data inconsistency has been minimized. However, it turns out that these artifacts can be reduced effectively by gray level segmentation. In this trivial example, we simply binarize the reconstruction result every $50$ iterations, and a perfect result can be obtained (Fig.\ref{example}h). It is worth mentioning that segmentation may not always increase reconstruction accuracy. Artifacts, blurred boundaries, and the detail of imaged objects may be incorrectly removed or enhanced. The key is that the incorrect perturbation caused by the segmentation can be corrected by conventional IR algorithms in the later iterations, while the artifacts can only be removed by the segmentation. An example is shown in Fig.\ref{example}c, where the artifacts far from the object have been removed correctly, but the artifacts close to the object have been incorrectly enhanced due to their high gray levels. However, in later iterations, the incorrectly enhanced artifacts are gradually corrected by the conventional IR algorithm (Fig.\ref{example}d). This process can be repeated multiple times (Fig.\ref{example}c to Fig.\ref{example}d, Fig.\ref{example}e to Fig.\ref{example}f), and a high-quality reconstruction result (Fig.\ref{example}h) can be finally obtained.

It is worth mentioning that the proposed constraint does not utilize any local information, so collaborations with most existing constraints are also possible. The pseudo-code for our method collaborating with these constraints is shown in Algorithm \ref{alg2}. Line $2$ to line $4$ depict the conventional ART algorithm with local constraints. The proposed method is described from line $5$ to line $19$. Line $5$ indicates that the proposed method will first execute at the $N_c$-th iteration since the correct segmentation needs a good initial reconstruction. This operation will be repeated every $N_c$ iterations, as the possible incorrect perturbation made by the proposed method needs to be corrected by conventional IR algorithms. In line $7$, pixels in the current result $\boldsymbol f$ will be clustered into $n$ groups. Otsu's method~\cite{otsu1979threshold} is used in this paper for simplicity. As for $n$, we propose to set $n = 3$ at the beginning, corresponding to the black area (air, etc.), bright area (bone, etc.), and the remaining area (body tissues, etc.). It is obvious that such a simple clustering cannot assist in generating detailed reconstruction results as the optimal $n$ is unknown and may vary for different objects. To overcome this challenge, $n$ is set to increase along the number of iterations ($n = \lfloor \frac{i}{N_c} \rfloor + 2$). In that case, when $n$ and $i$ are small, and the reconstruction is of low quality, the method focuses on removing the evident artifacts. When $n$ and $i$ are large, and the reconstruction is of high quality, the method aims for more detailed segmentation and artifact removal. It is worth mentioning that the proposed method is still effective even if $n$ is larger than the actual number of tissue types in the imaged object. This is because pixels corresponding to the same tissue may have slightly different gray levels. In that case, it is more reasonable to let $n$ be larger than the number of tissue types. Ideally, the proposed method can be repeatedly executed until $n$ equals to the number of pixels. However, the proposed method will terminate after $N_{stop}$-th iteration. The reason is the decreasing improvement and the increasing complexity of the proposed method for large $n$. The approximated segmentation will be refined at line $8$ to line $10$, where a pixel will be removed from its group if not all its neighbors belong to the same group. After that, the median of each group's gray level is calculated and assigned to all the pixels in that group from line $11$ to line $16$. As a result, $\boldsymbol f_{seg}$ is obtained and used for modifying the reconstruction result $\boldsymbol f$ in line $17$ and line $18$. $\beta \in [0,1]$ is the step size of the update. It can be set to $1$ if we believe that the imaged object is piece-wise constant.

\begin{algorithm}[htb]
    \caption{Reconstruction Using Both Local and Global Priors}
    \label{alg2}
    \hspace*{0.02in} {\bf Parameters:} 
    $\alpha, \beta, N_g, N_c, N_{stop},N_{iter}$
    \begin{algorithmic}[1]
        \State {\bf for} $i = 1,2,3, ..., N_{iter}$ {\bf do} 
        \State \quad $\boldsymbol f$ = ART loop with step size $\alpha$
        \State \quad $\boldsymbol f$ = enforce positivity
        \State \quad local constraint (e.g. TV, ATV, qGGMRF) minimization with step size $\beta$ ($N_g$ times)
        \State \quad {\bf if} $i \bmod N_c = 0$ and $i < N_{stop}$ {\bf then}:
        \State \quad \quad $\boldsymbol f_{seg} = \boldsymbol f$
        \State \quad \quad cluster the pixels in $\boldsymbol f$ into $n = \lfloor \frac{i}{N_c} \rfloor + 2$ groups based on their gray levels
        \State \quad \quad {\bf for} every pixel:
        \State \quad \quad \quad remove it from its group {\bf if not} all its neighbors belong to the same group
        \State \quad \quad {\bf end for}
        \State \quad \quad {\bf for} every group:
        \State \quad \quad \quad calculate the median of its gray levels
        \State \quad \quad {\bf end for}
        \State \quad \quad {\bf for} every pixel in $\boldsymbol f_{seg}$:
        \State \quad \quad \quad set its gray value to the corresponding group median if it belongs to a group
        \State \quad \quad {\bf end for}
        \State \quad \quad $\boldsymbol d = \boldsymbol f-\boldsymbol f_{seg}$
        \State \quad \quad $\boldsymbol f = \boldsymbol f- \beta \boldsymbol d$
        \State \quad {\bf end if}
        \State {\bf end for}
        \State {\bf return} $\boldsymbol f$
    \end{algorithmic}
\end{algorithm}

\section{Experiments and Results}
\label{sec:exp}
In this section, a series of experiments are carried out to evaluate the improvement made by the proposed method based on TV, RwATV, and qGGMRF, all state-of-the-art constraints. To demonstrate the effectiveness of the proposed method on not only phantom but also different real objects, the proposed method is tested on three images (first columns of Fig.\ref{resultsp} and Fig.\ref{resultla}), the first one is the Shepp-Logan Phantom ($512 \times 512$)~\cite{shepp1974fourier}, the second one is a real CT image ($512 \times 512$) of the human brain from HNSCC-3DCT-RT dataset~\cite{bejarano9head}, the last one is a real CT image ($362 \times 362$) of the human lung from LoDoPaB-CT dataset~\cite{leuschner2021lodopab}. In our experiments, the parameters are set to $\alpha = 0.2, \beta = 0.5, N_g = 20, N_c = 50, N_{stop}=800,$ and $ N_{iter} = 1000$. The parameters for RwATV are: $a = 1, b = 0.001$; the parameters for qGGMRF are: $p=2,q=1,$ and $c=0.0625$. These parameters are optimized according to all three aforementioned images corresponding to different objects under multiple conditions (sparse-view, limited-angle, and low-dose), higher performance is obtainable if these parameters are optimized for each individual case.

\subsection{The Effect of the Proposed Method}\label{sec:3.1}
\begin{figure}[pos=htb]
    \begin{minipage}[b]{0.32\linewidth}
        \centering
        \centerline{\includegraphics[width=3cm]{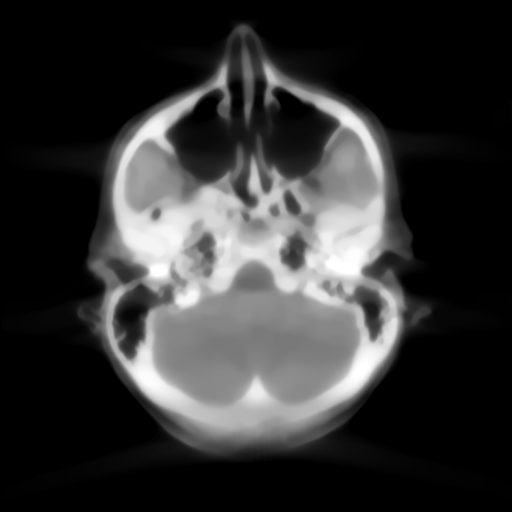}}
    \end{minipage}
    \begin{minipage}[b]{0.32\linewidth}
        \centering
        \centerline{\includegraphics[width=3cm]{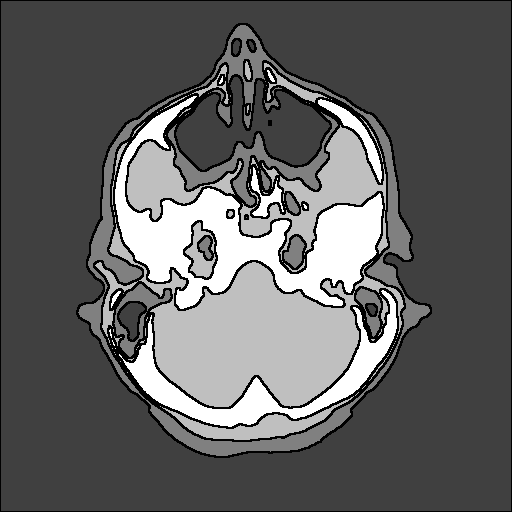}}
    \end{minipage}
    \begin{minipage}[b]{0.32\linewidth}
        \centering
        \centerline{\includegraphics[width=3cm]{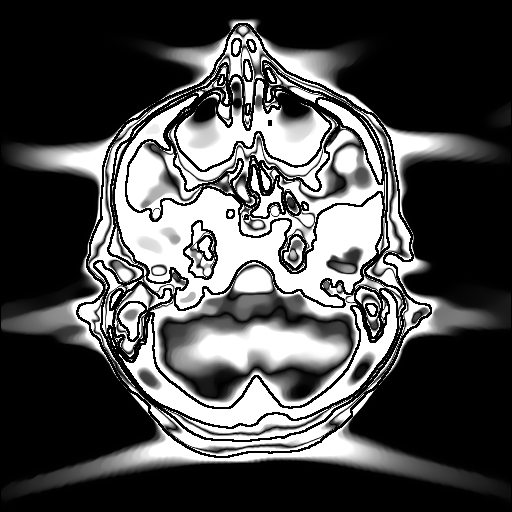}}
    \end{minipage}
    \centerline{(a) 100th iteration, $n$ = 4}\medskip
    
    \begin{minipage}[b]{0.32\linewidth}
        \centering
        \centerline{\includegraphics[width=3cm]{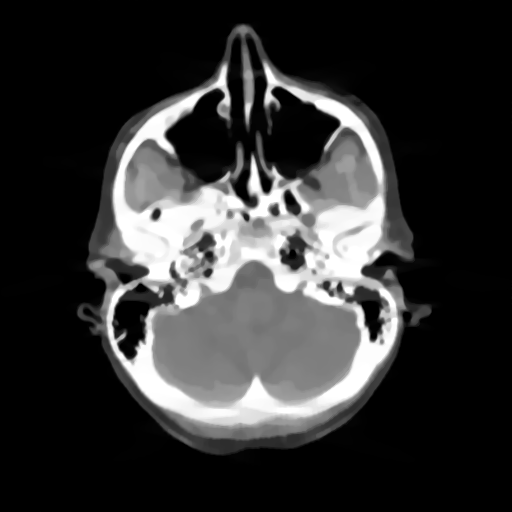}}
    \end{minipage}
    \begin{minipage}[b]{0.32\linewidth}
        \centering
        \centerline{\includegraphics[width=3cm]{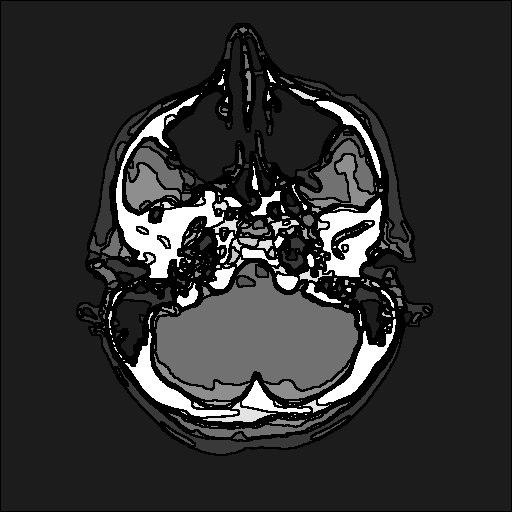}}
    \end{minipage}
    \begin{minipage}[b]{0.32\linewidth}
        \centering
        \centerline{\includegraphics[width=3cm]{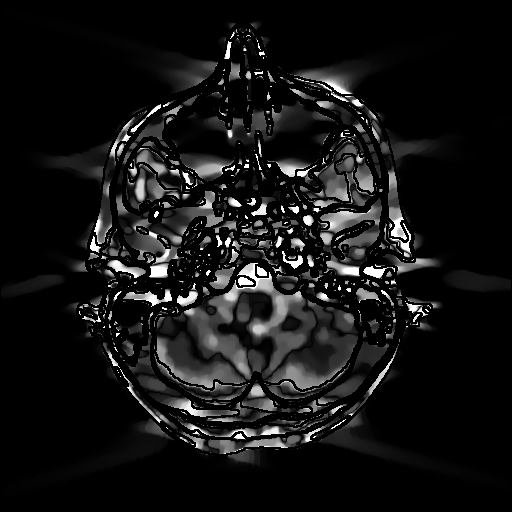}}
    \end{minipage}
    \centerline{(b) 300th iteration, $n=8$}\medskip
    
    \begin{minipage}[b]{0.32\linewidth}
        \centering
        \centerline{\includegraphics[width=3cm]{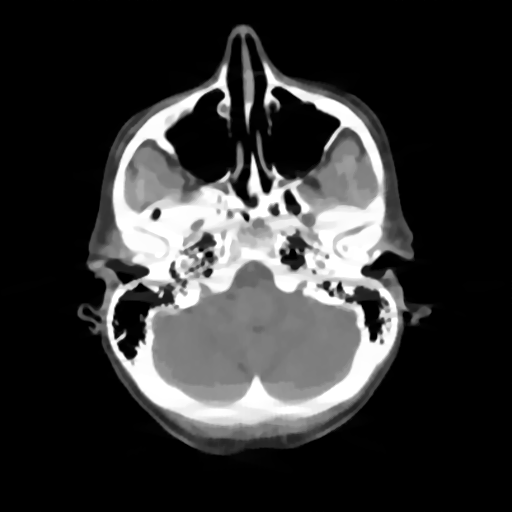}}
    \end{minipage}
    \begin{minipage}[b]{0.32\linewidth}
        \centering
        \centerline{\includegraphics[width=3cm]{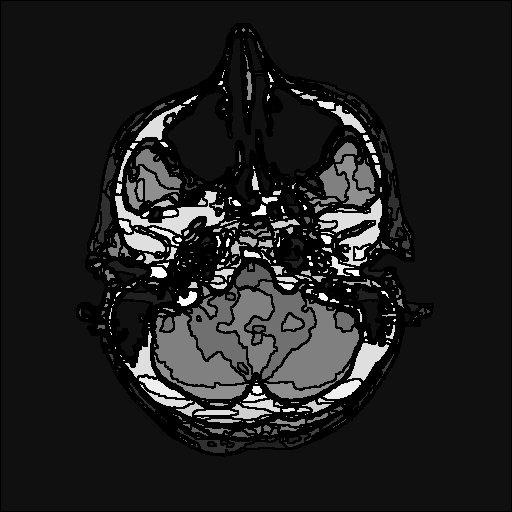}}
    \end{minipage}
    \begin{minipage}[b]{0.32\linewidth}
        \centering
        \centerline{\includegraphics[width=3cm]{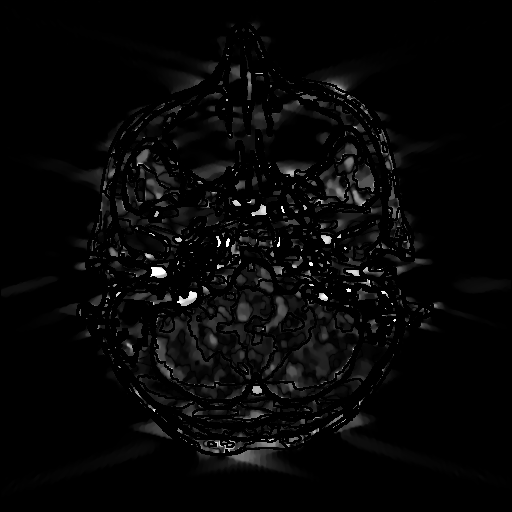}}
    \end{minipage}
    \centerline{(c) 700th iteration, $n=16$}\medskip
    \caption{The effect of the proposed method at the different number of iterations. The first column shows the reconstruction before the proposed method. The second column shows the segmentation generated by the proposed method, where different gray levels indicate different groups. The third column shows the difference between the first column and the segmentation ($\boldsymbol d$ in Algorithm \ref{alg2}), the gray level window is set to $[0,0.05]$}
    \label{reconprocess}
\end{figure}

\begin{figure}[pos=h]
    \begin{minipage}[b]{0.48\linewidth}
    \centering
    \centerline{\includegraphics[width=4.3cm]{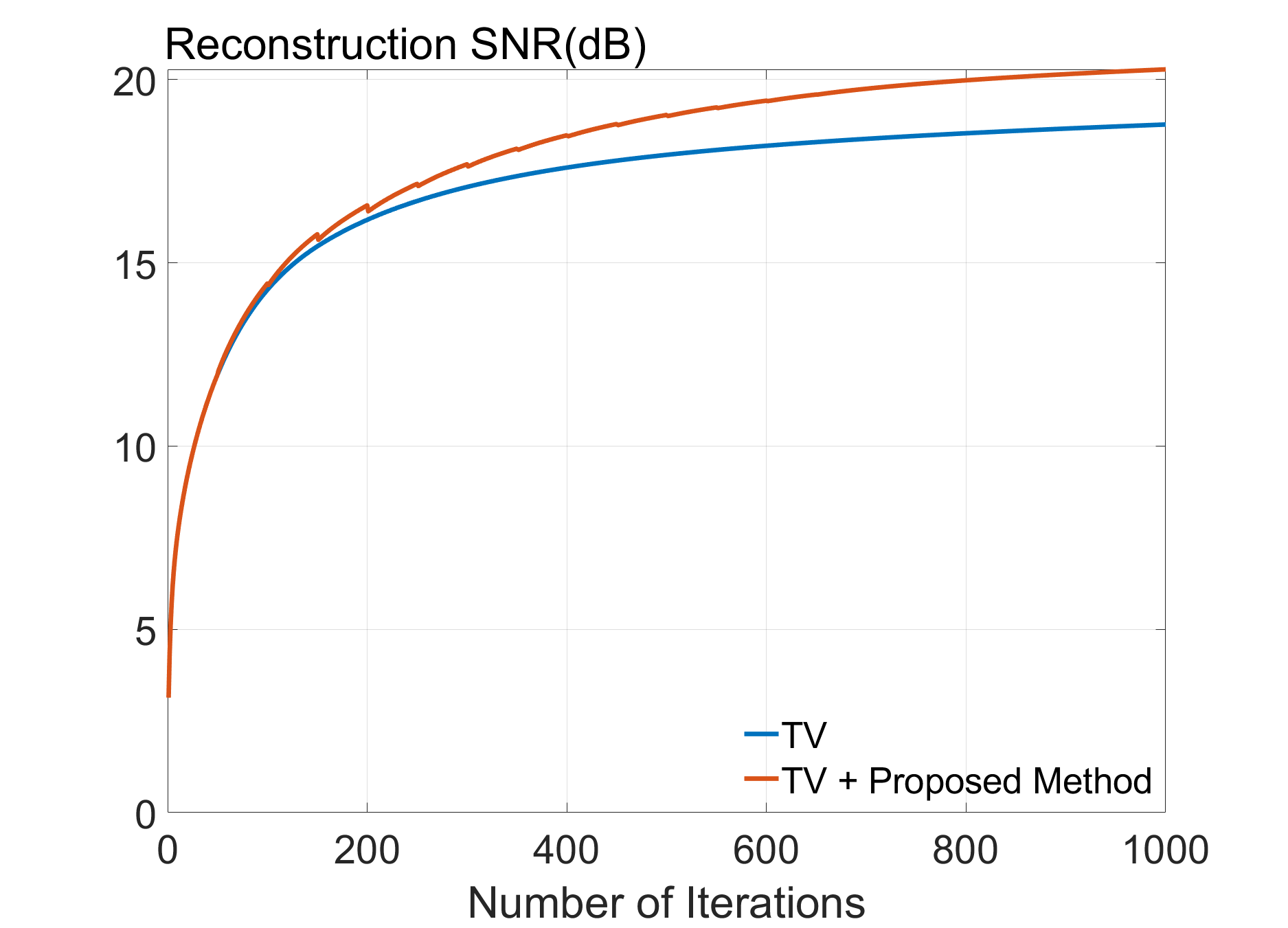}}
    \centerline{(a)}\medskip
\end{minipage}
\begin{minipage}[b]{0.48\linewidth}
    \centering
    \centerline{\includegraphics[width=4.3cm]{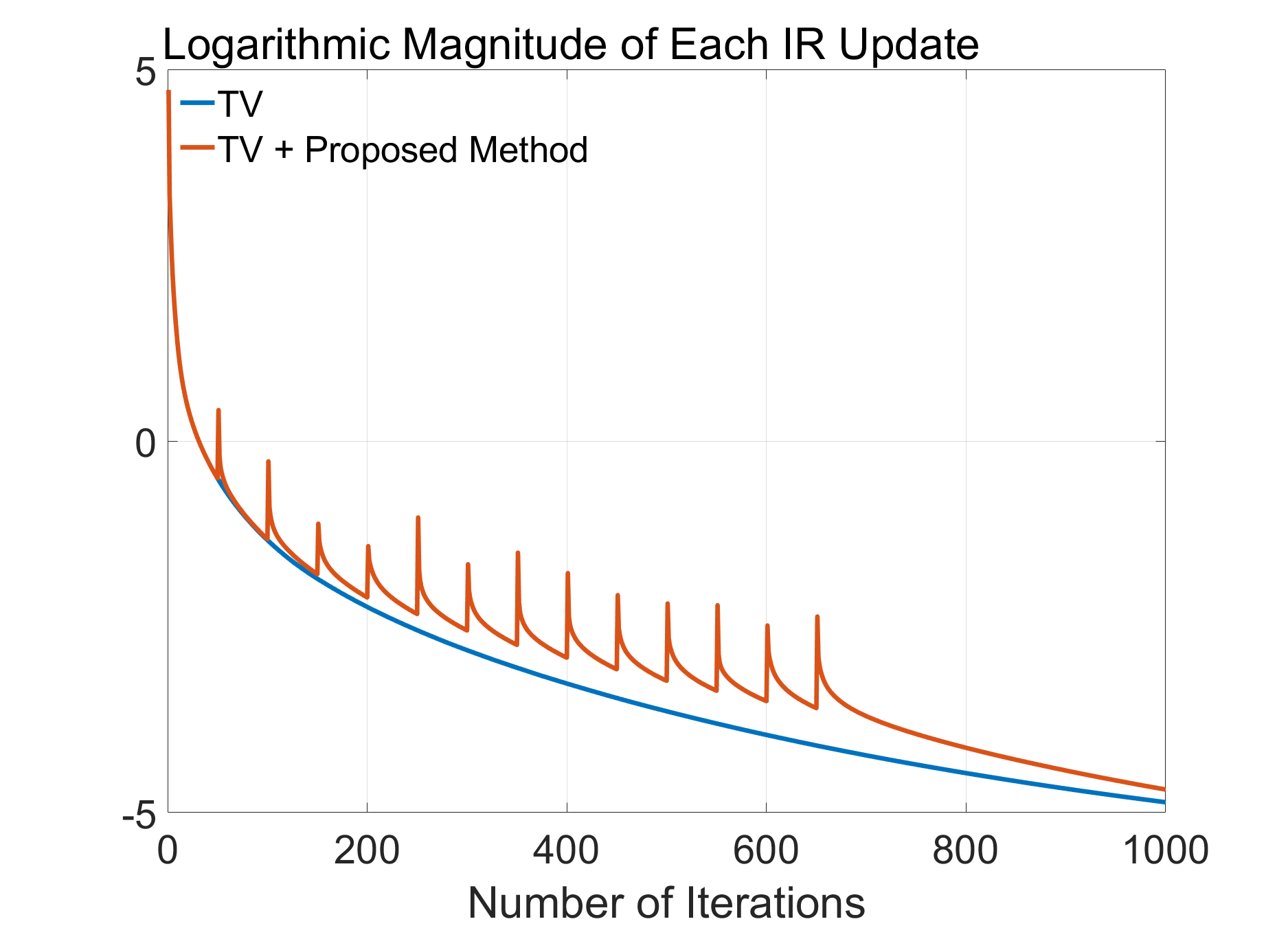}}
    \centerline{(b)}\medskip
\end{minipage}
\caption{(a) The SNR and (b) the magnitude of update at different numbers of iterations for the reconstruction in Section \ref{sec:3.1}.}
\label{reconprocess2}
\end{figure}

In this section, a limited-angle ($15^\circ$ to $165^\circ$, one view per degree) CT reconstruction for the human brain is presented as an example to illustrate the actual effect of the proposed method in IR algorithms. As shown in Fig.\ref{reconprocess}a, at the early stage of the reconstruction, where $n$ is relatively small, the proposed method focuses on an estimated segmentation to classify different tissues. This can be utilized to remove artifacts effectively. However, pixels belonging to the same tissue may still have different gray levels in practice. Thus, $n$ is set to increase along with the number of iterations. In Fig.\ref{reconprocess}b, pixels are grouped into $8$ groups for a more detailed artifact removal. As the number of iterations increases, $n$ may be larger than optimal. As shown in Fig.\ref{reconprocess}c, some of the pixels are improperly clustered into superfluous groups. Even so, it will not noticeably downgrade the reconstruction accuracy. The reason is that the proposed method enforces the same gray level (group median) for the pixels in the same group. Increasing the number of groups will decrease the absolute deviations around their medians. Thus, the perturbation caused by the proposed method will also decrease. The SNR and the magnitude of update at different numbers of iterations are illustrated in Fig.\ref{reconprocess2}. Because of the proposed method, the SNR decreases slightly and then resumes its upward trend every $50$ iteration (after executing the proposed method), resulting in a more accurate reconstruction.

\subsection{Sparse-View and Limited-Angle CT Reconstruction}
\begin{figure*}
    \centering
    \begin{minipage}[b]{.14\linewidth}
        \centering
        \centerline{\includegraphics[width=\linewidth]{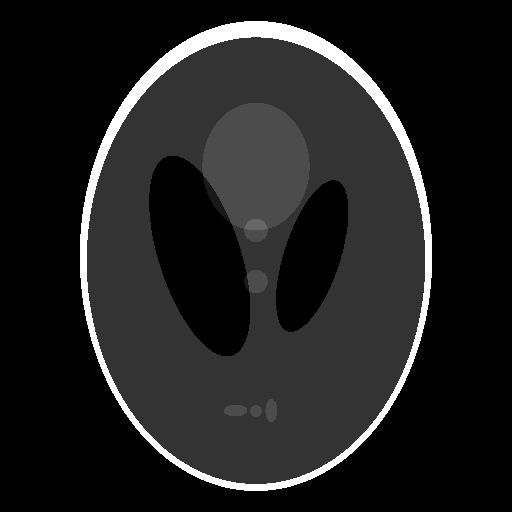}}
        \centerline{ }\medskip
    \end{minipage}
    \begin{minipage}[b]{.14\linewidth}
        \centering
        \centerline{\includegraphics[width=\linewidth]{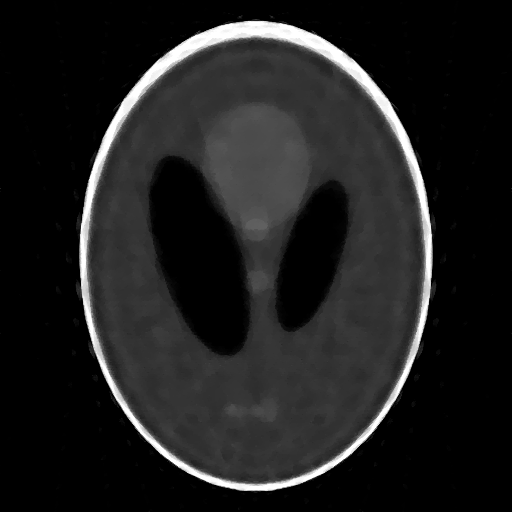}}
        \centerline{14.22dB}\medskip
    \end{minipage}
    \begin{minipage}[b]{.14\linewidth}
        \centering
        \centerline{\includegraphics[width=\linewidth]{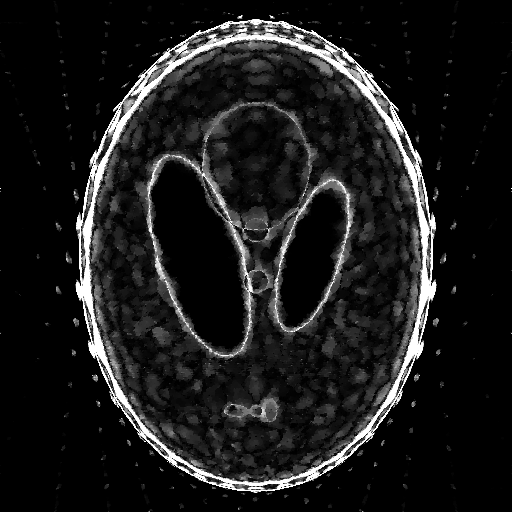}}
        \centerline{}\medskip
    \end{minipage}
    \begin{minipage}[b]{.14\linewidth}
        \centering
        \centerline{\includegraphics[width=\linewidth]{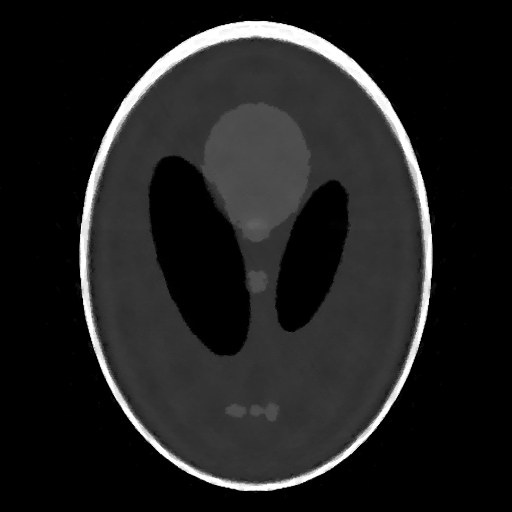}}
        \centerline{15.81dB}\medskip
    \end{minipage}
    \begin{minipage}[b]{.14\linewidth}
        \centering
        \centerline{\includegraphics[width=\linewidth]{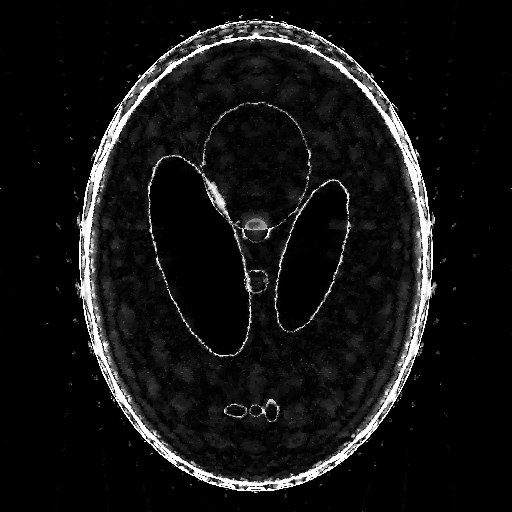}}
        \centerline{}\medskip
    \end{minipage}
    \vspace{-0.2cm}

    \begin{minipage}[b]{.14\linewidth}
        \centering
        \centerline{\includegraphics[width=\linewidth]{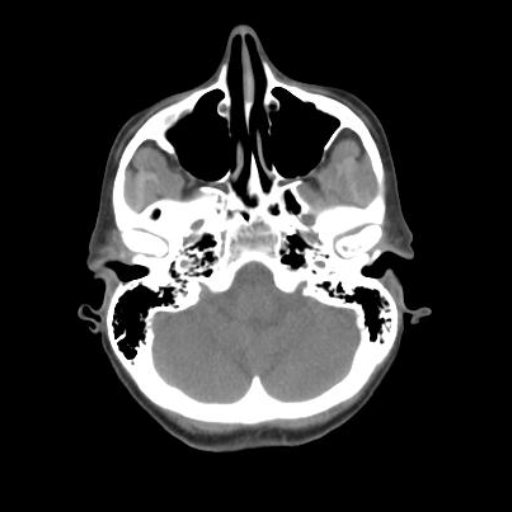}}
        \centerline{ }\medskip
    \end{minipage}
    \begin{minipage}[b]{.14\linewidth}
        \centering
        \centerline{\includegraphics[width=\linewidth]{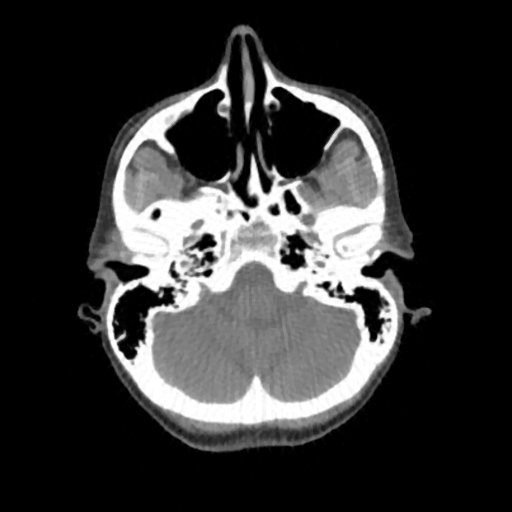}}
        \centerline{26.30dB}\medskip
    \end{minipage}
    \begin{minipage}[b]{.14\linewidth}
        \centering
        \centerline{\includegraphics[width=\linewidth]{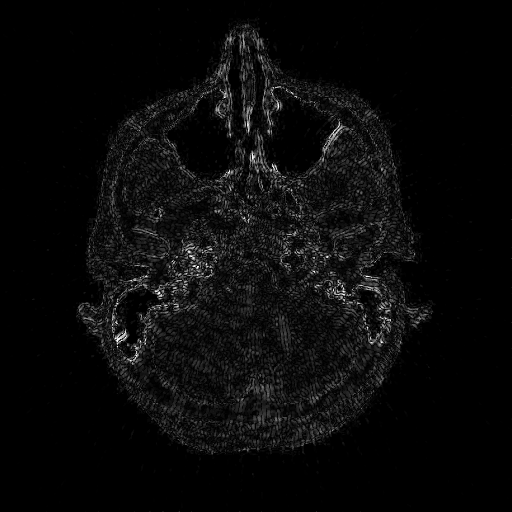}}
        \centerline{}\medskip
    \end{minipage}
    \begin{minipage}[b]{.14\linewidth}
        \centering
        \centerline{\includegraphics[width=\linewidth]{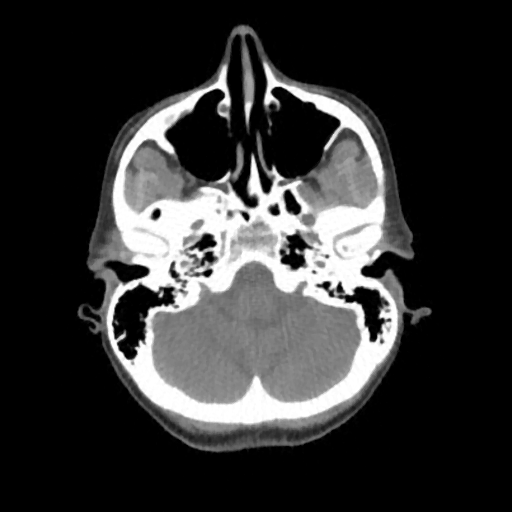}}
        \centerline{28.49dB}\medskip
    \end{minipage}
    \begin{minipage}[b]{.14\linewidth}
        \centering
        \centerline{\includegraphics[width=\linewidth]{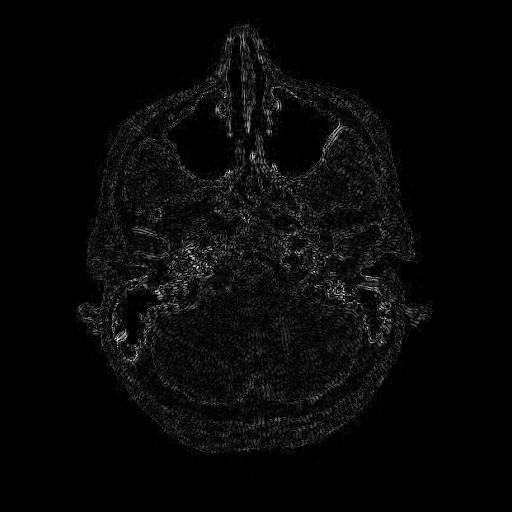}}
        \centerline{}\medskip
    \end{minipage}
    \vspace{-0.2cm}

    \begin{minipage}[b]{.14\linewidth}
        \centering
        \centerline{\includegraphics[width=\linewidth]{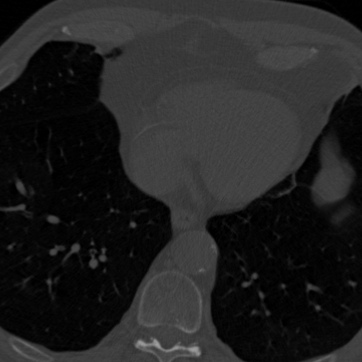}}
        \centerline{ }\medskip
        \centerline{(a)}\medskip
    \end{minipage}
    \begin{minipage}[b]{.14\linewidth}
        \centering
        \centerline{\includegraphics[width=\linewidth]{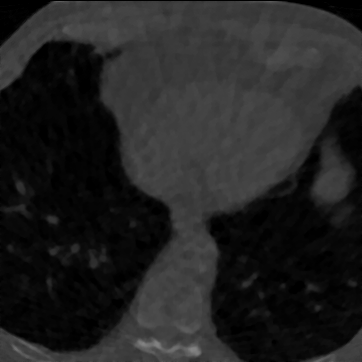}}
        \centerline{18.23dB}\medskip
        \centerline{(b)}\medskip
    \end{minipage}
    \begin{minipage}[b]{.14\linewidth}
        \centering
        \centerline{\includegraphics[width=\linewidth]{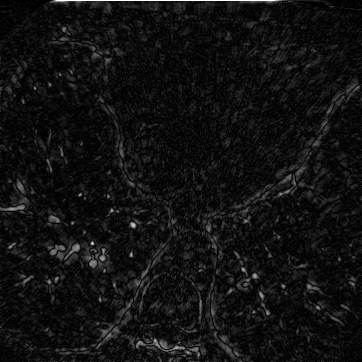}}
        \centerline{}\medskip
        \centerline{(c)}\medskip
    \end{minipage}
    \begin{minipage}[b]{.14\linewidth}
        \centering
        \centerline{\includegraphics[width=\linewidth]{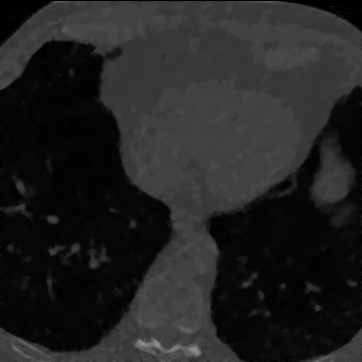}}
        \centerline{19.94dB}\medskip
        \centerline{(d)}\medskip
    \end{minipage}
    \begin{minipage}[b]{.14\linewidth}
        \centering
        \centerline{\includegraphics[width=\linewidth]{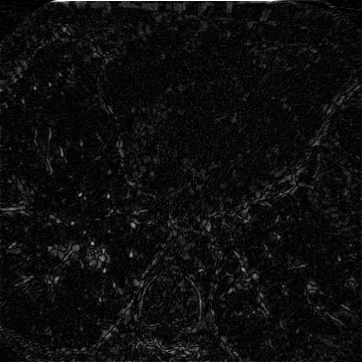}}
        \centerline{}\medskip
        \centerline{(e)}\medskip
    \end{minipage}
    \vspace{-0.2cm}
    \caption{The sparse-view reconstruction results of the Shepp-Logan phantom (15 views, first row), a brain CT from HNSCC-3DCT-RT dataset (90 views, second row), and a lung CT from LoDoPaB dataset (30 views, third row) with and without the proposed method. Column (a) shows the ground truth, column (b) shows the result using TV constraint, column (c) shows the absolute error between (a) and (b), column (d) shows the result using TV and the proposed method, column (e) shows the absolute error between (a) and (d). The gray level windows of columns (c) and (e) are set to $[0,0.05]$}
    \label{resultsp}
\end{figure*}

\begin{figure*}
    \centering
    \begin{minipage}[b]{.14\linewidth}
        \centering
        \centerline{\includegraphics[width=\linewidth]{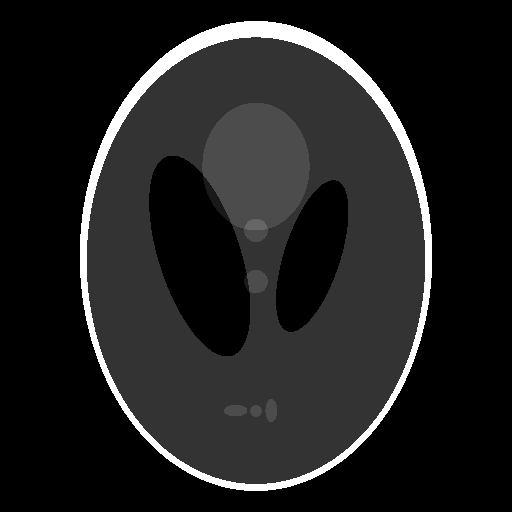}}
        \centerline{ }\medskip
    \end{minipage}
    \begin{minipage}[b]{.14\linewidth}
        \centering
        \centerline{\includegraphics[width=\linewidth]{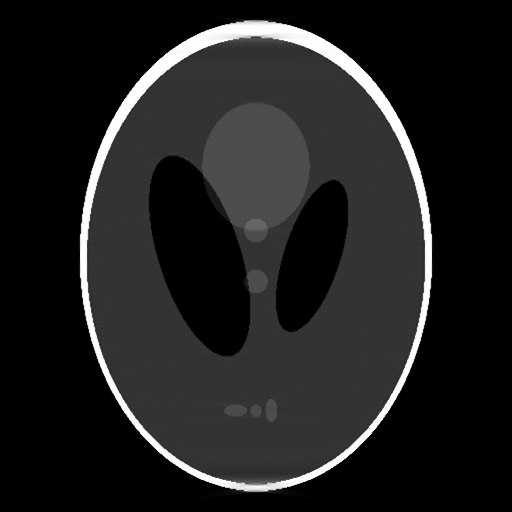}}
        \centerline{20.92dB}\medskip
    \end{minipage}
    \begin{minipage}[b]{.14\linewidth}
        \centering
        \centerline{\includegraphics[width=\linewidth]{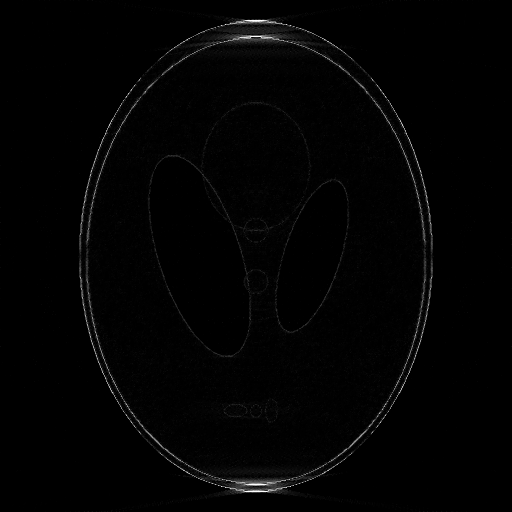}}
        \centerline{}\medskip
    \end{minipage}
    \begin{minipage}[b]{.14\linewidth}
        \centering
        \centerline{\includegraphics[width=\linewidth]{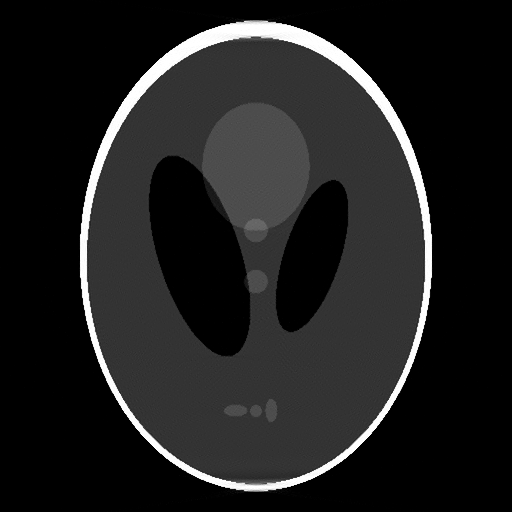}}
        \centerline{23.72dB}\medskip
    \end{minipage}
    \begin{minipage}[b]{.14\linewidth}
        \centering
        \centerline{\includegraphics[width=\linewidth]{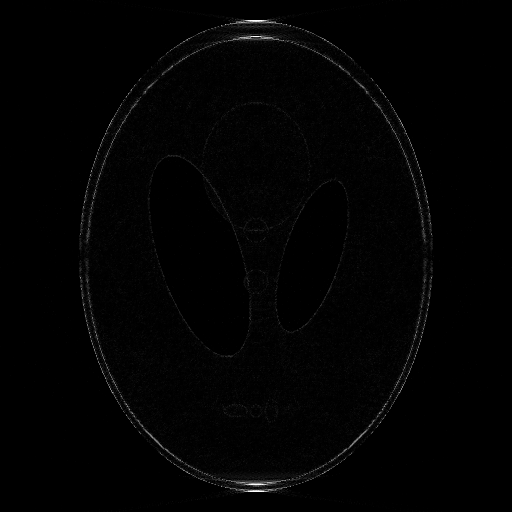}}
        \centerline{}\medskip
    \end{minipage}

    \vspace{-0.2cm}
    \begin{minipage}[b]{.14\linewidth}
        \centering
        \centerline{\includegraphics[width=\linewidth]{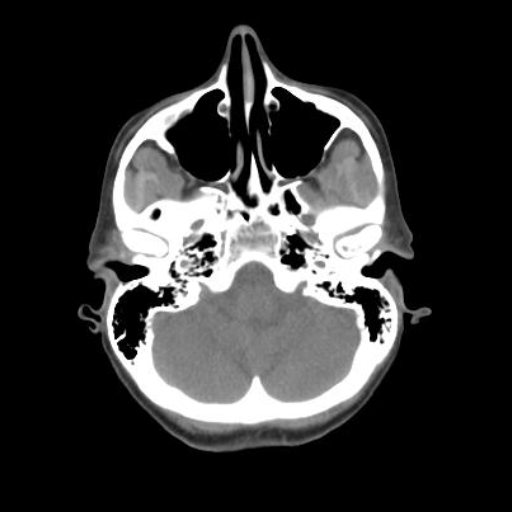}}
        \centerline{}\medskip
    \end{minipage}
    \begin{minipage}[b]{.14\linewidth}
        \centering
        \centerline{\includegraphics[width=\linewidth]{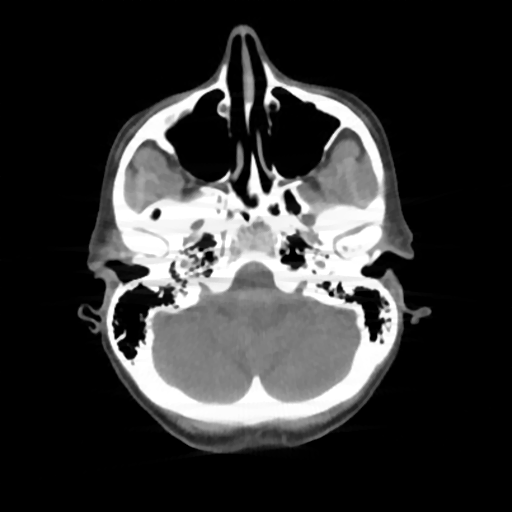}}
        \centerline{24.44dB}\medskip
    \end{minipage}
    \begin{minipage}[b]{.14\linewidth}
        \centering
        \centerline{\includegraphics[width=\linewidth]{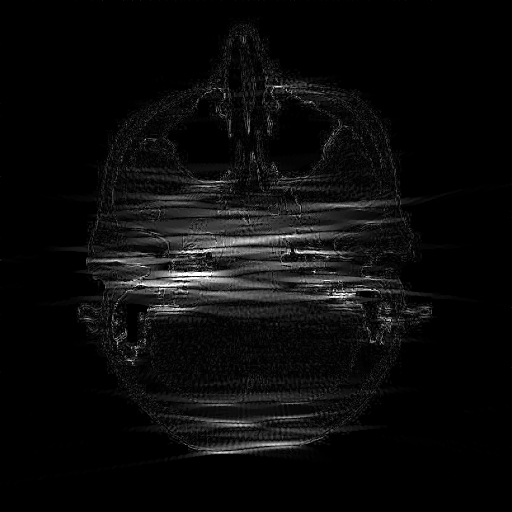}}
        \centerline{}\medskip
    \end{minipage}
    \begin{minipage}[b]{.14\linewidth}
        \centering
        \centerline{\includegraphics[width=\linewidth]{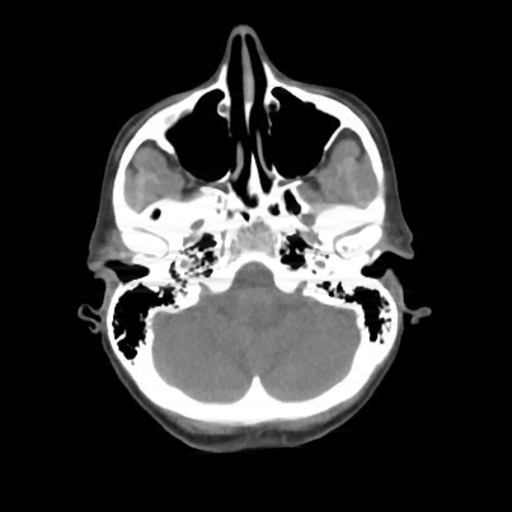}}
        \centerline{27.96dB}\medskip
    \end{minipage}
    \begin{minipage}[b]{.14\linewidth}
        \centering
        \centerline{\includegraphics[width=\linewidth]{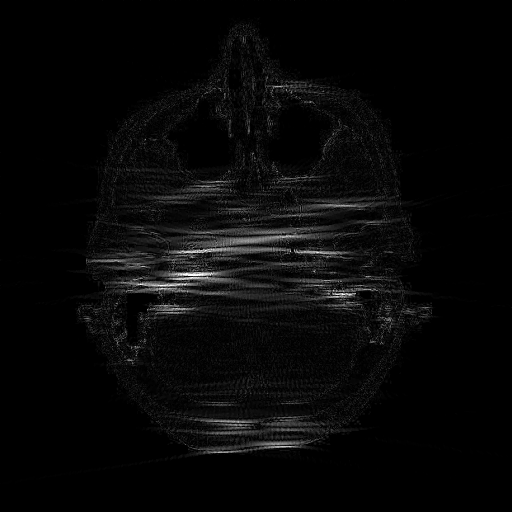}}
        \centerline{}\medskip
    \end{minipage}
    \vspace{-0.2cm}

    \begin{minipage}[b]{.14\linewidth}
        \centering
        \centerline{\includegraphics[width=\linewidth]{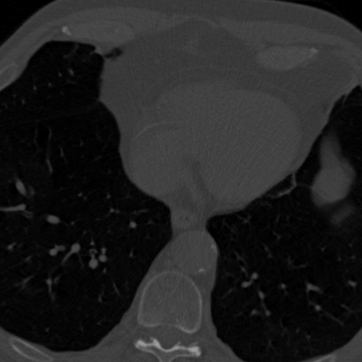}}
        \centerline{ }\medskip
        \centerline{(a)}\medskip
    \end{minipage}
    \begin{minipage}[b]{.14\linewidth}
        \centering
        \centerline{\includegraphics[width=\linewidth]{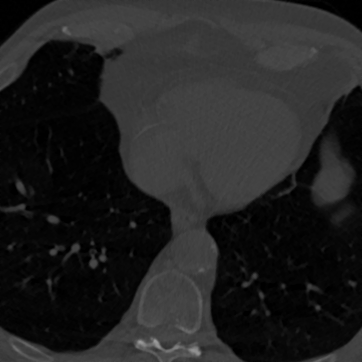}}
        \centerline{27.05dB}\medskip
        \centerline{(b)}\medskip
    \end{minipage}
    \begin{minipage}[b]{.14\linewidth}
        \centering
        \centerline{\includegraphics[width=\linewidth]{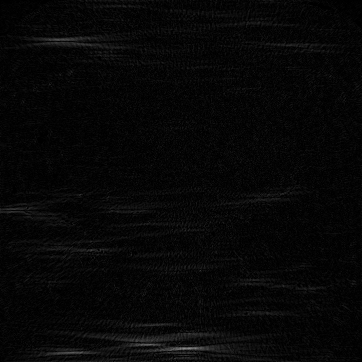}}
        \centerline{}\medskip
        \centerline{(c)}\medskip
    \end{minipage}
    \begin{minipage}[b]{.14\linewidth}
        \centering
        \centerline{\includegraphics[width=\linewidth]{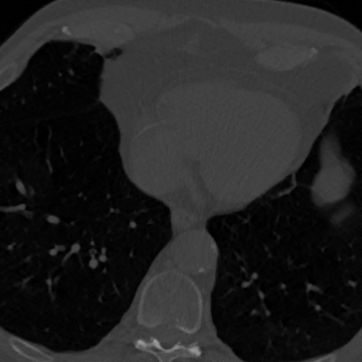}}
        \centerline{28.93dB}\medskip
        \centerline{(d)}\medskip
    \end{minipage}
    \begin{minipage}[b]{.14\linewidth}
        \centering
        \centerline{\includegraphics[width=\linewidth]{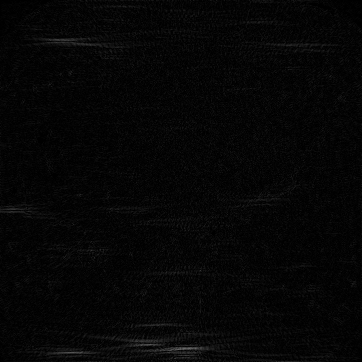}}
        \centerline{}\medskip
        \centerline{(e)}\medskip
    \end{minipage}
    \vspace{-0.2cm}
    \caption{The limited-angle ($15^\circ$ to $165^\circ$, one view per degree) reconstruction results of the Shepp-Logan phantom, a brain CT from HNSCC-3DCT-RT dataset, and a lung CT from LoDoPaB dataset with and without the proposed method. Column (a) shows the ground truth, column (b) shows the result using TV constraint, column (c) shows the absolute error between (a) and (b), column (d) shows the result using TV and the proposed method, column (e) shows the absolute error between (a) and (d). The gray level windows of columns (c) and (e) are set to $[0,0.05]$}
    \label{resultla}
\end{figure*}

\begin{figure}[pos=htb]
    \begin{minipage}[b]{0.32\linewidth}
        \centering
        \centerline{\includegraphics[width=3cm]{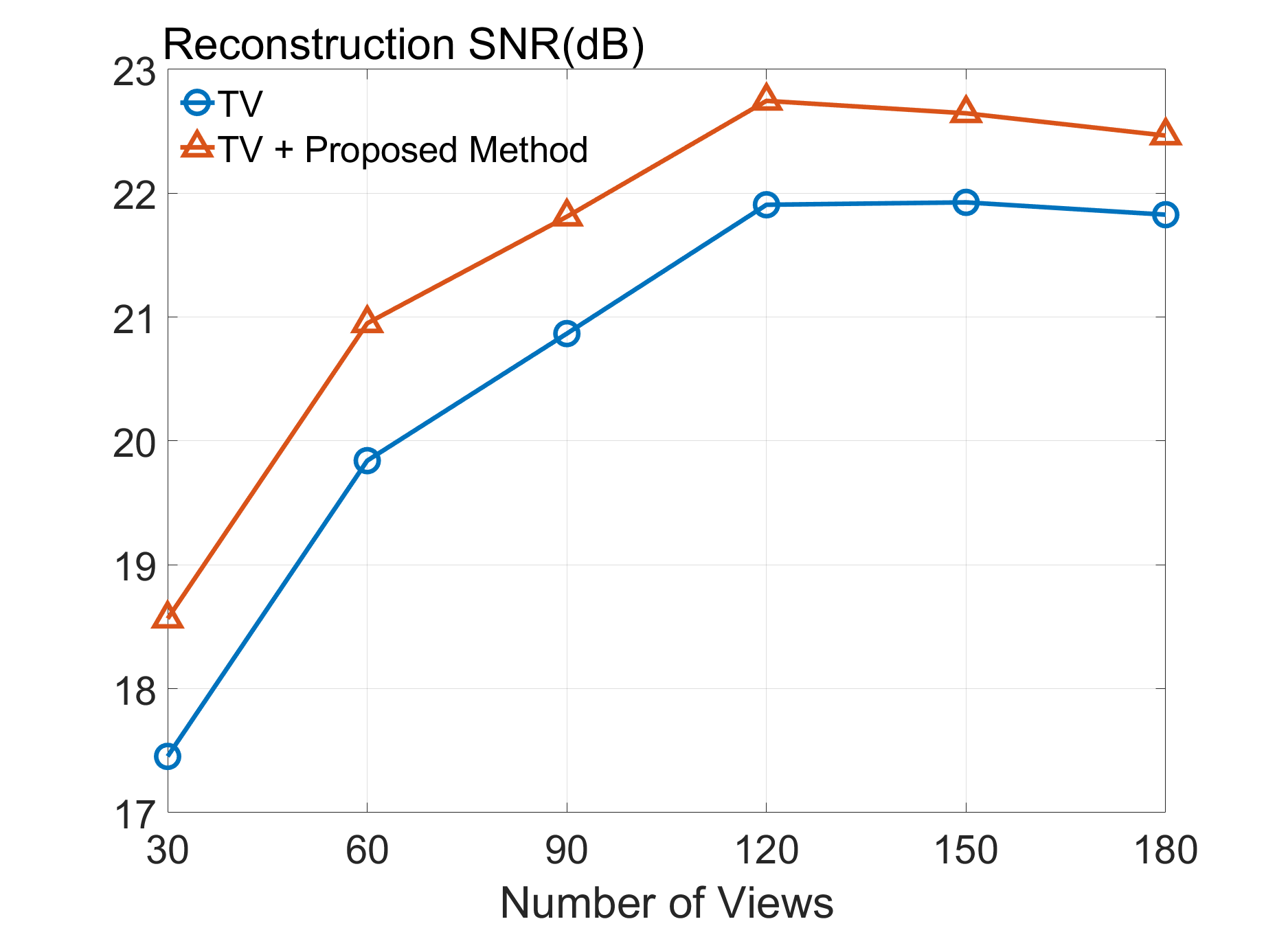}}
    \end{minipage}
    \begin{minipage}[b]{0.32\linewidth}
        \centering
        \centerline{\includegraphics[width=3cm]{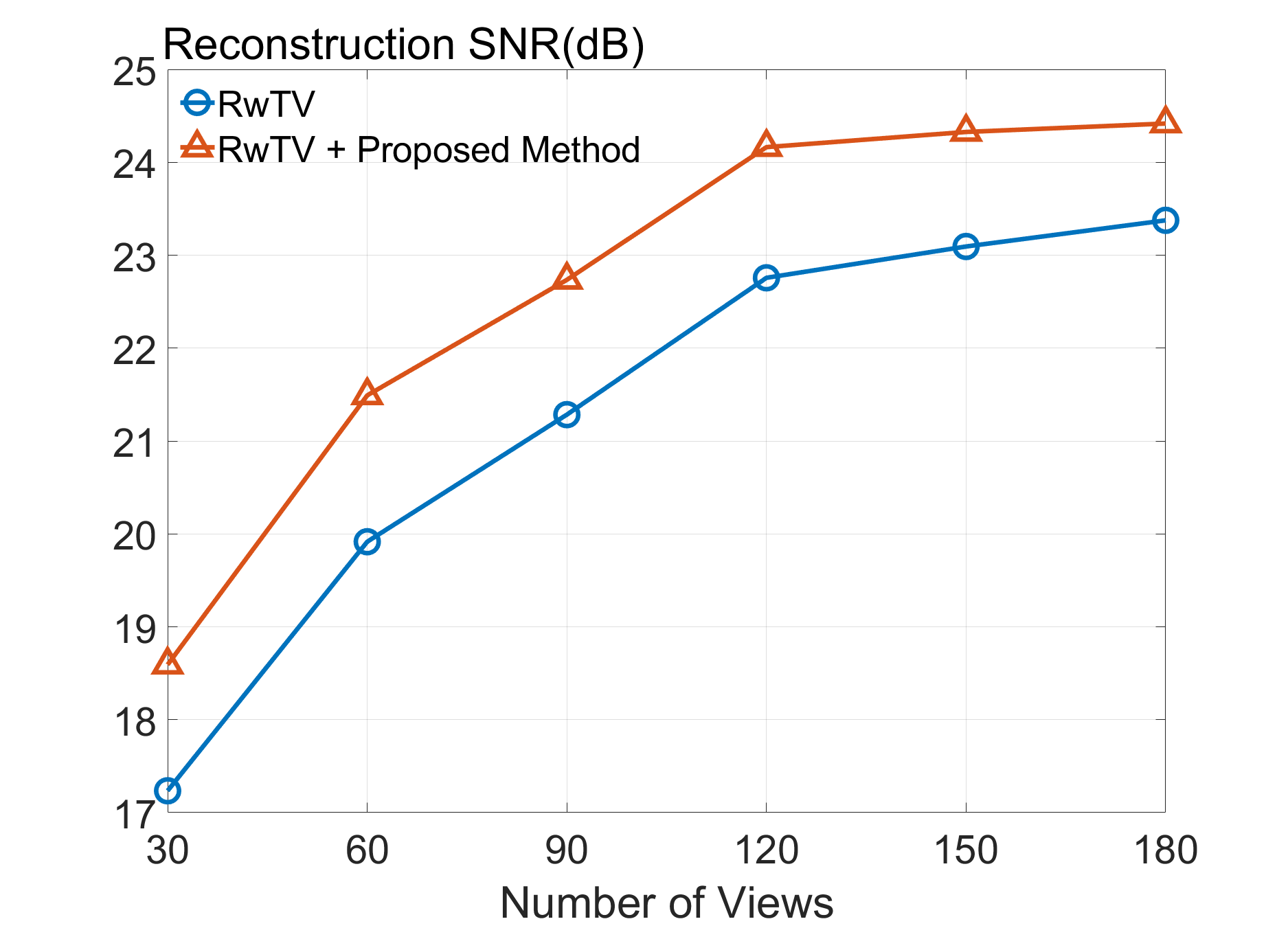}}
    \end{minipage}
    \begin{minipage}[b]{0.32\linewidth}
        \centering
        \centerline{\includegraphics[width=3cm]{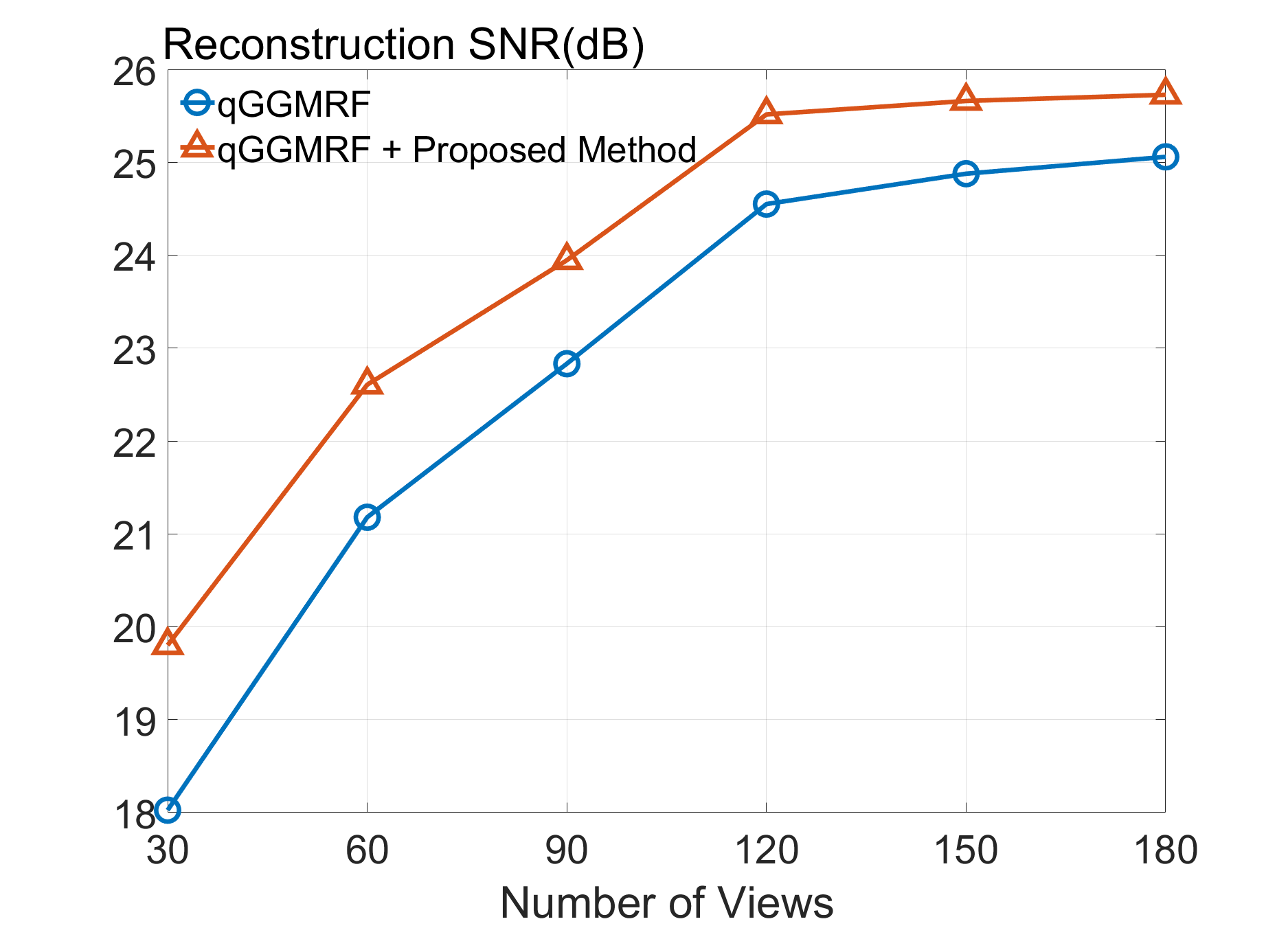}}
    \end{minipage}
    
    \begin{minipage}[b]{0.32\linewidth}
        \centering
        \centerline{\includegraphics[width=3cm]{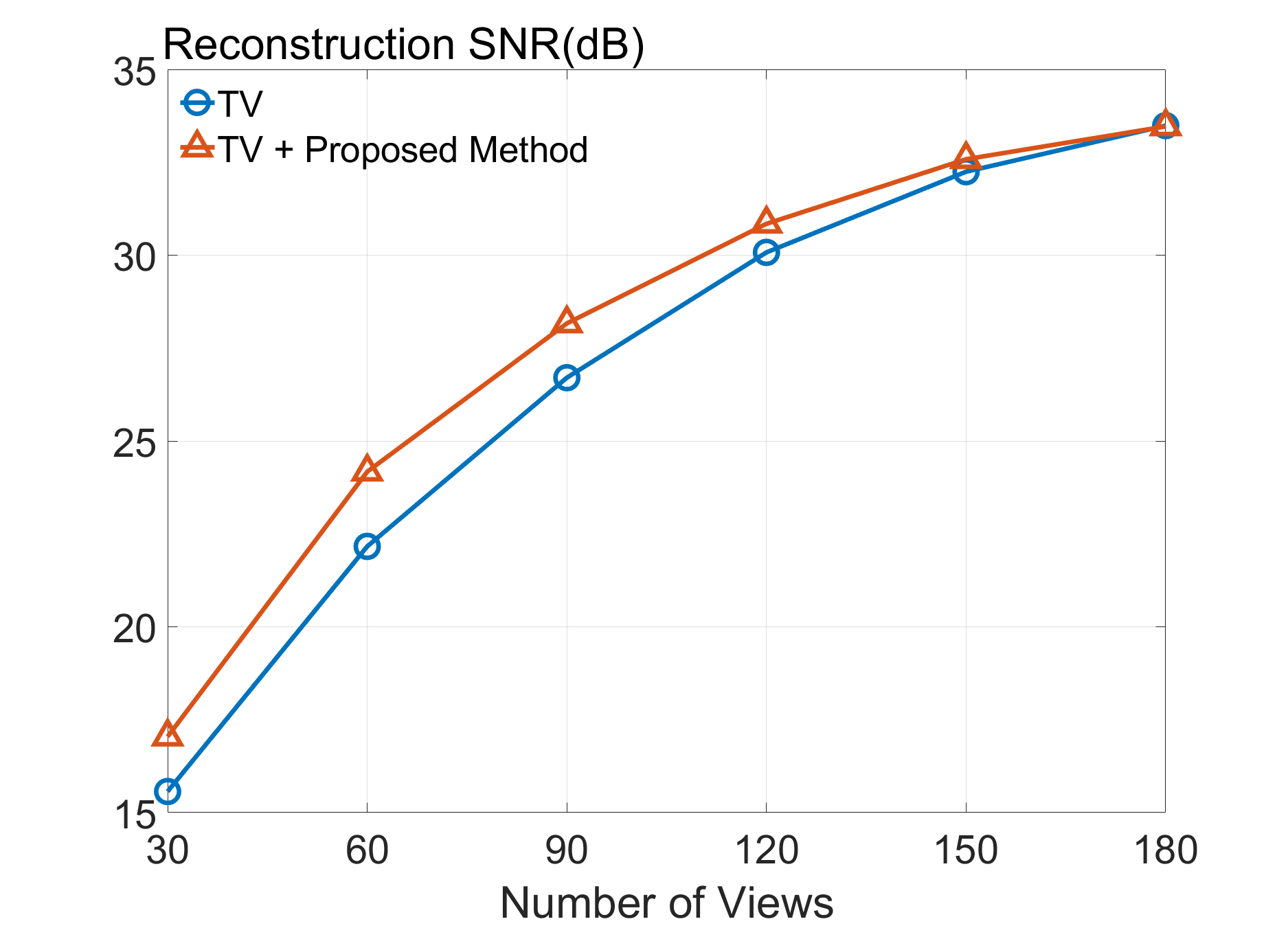}}
    \end{minipage}
    \begin{minipage}[b]{0.32\linewidth}
        \centering
        \centerline{\includegraphics[width=3cm]{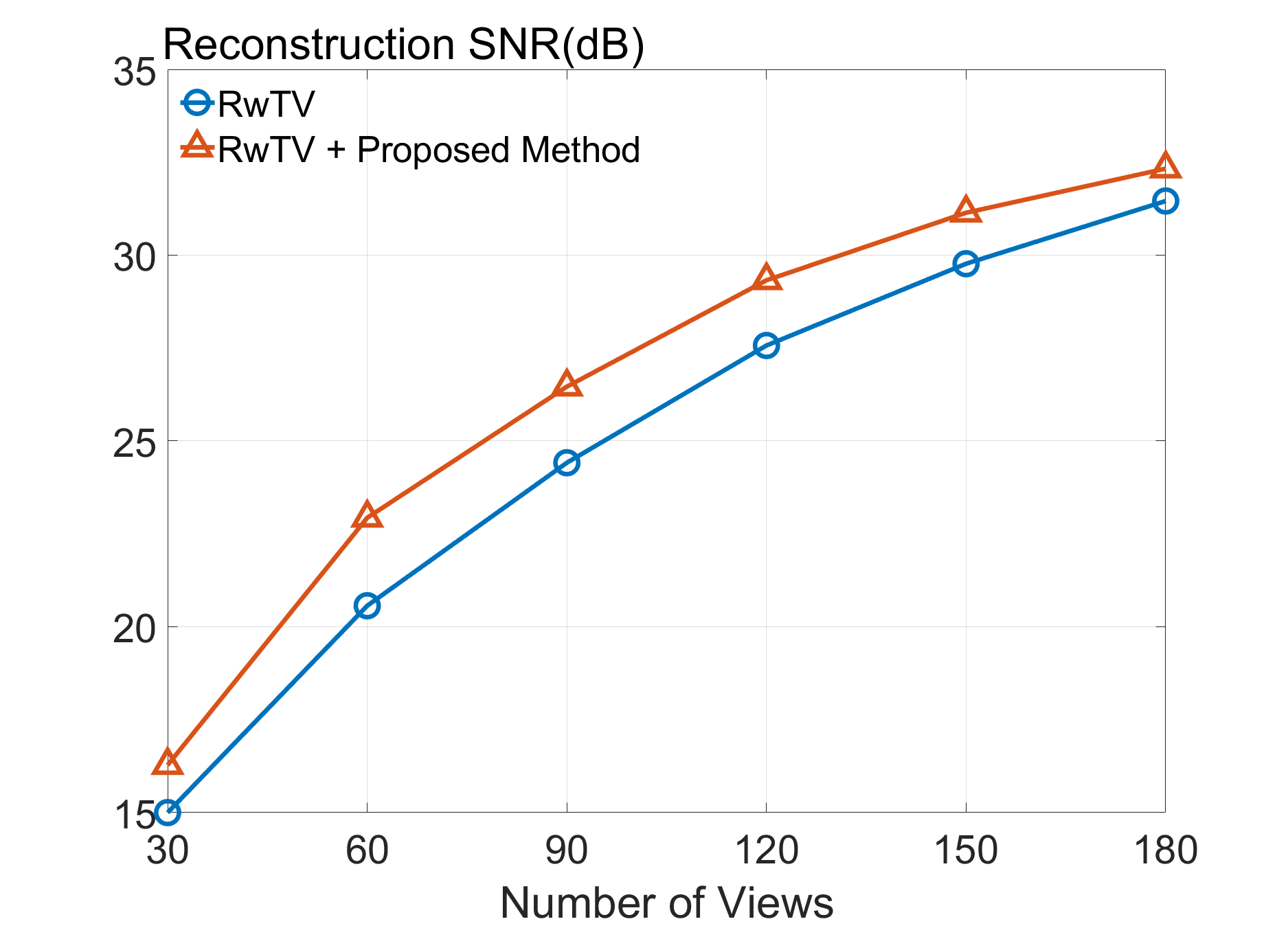}}
    \end{minipage}
    \begin{minipage}[b]{0.32\linewidth}
        \centering
        \centerline{\includegraphics[width=3cm]{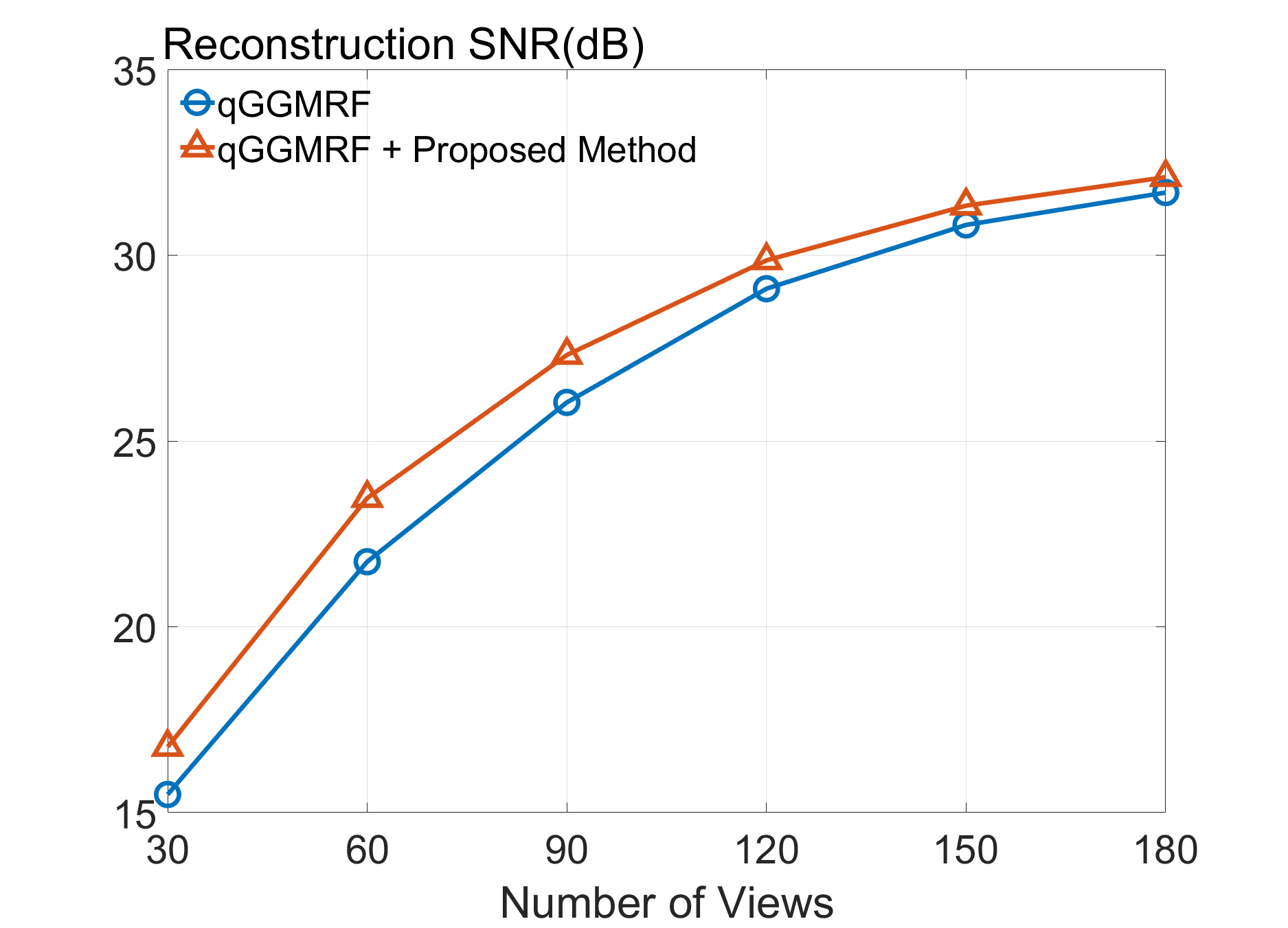}}
    \end{minipage}
    
    \begin{minipage}[b]{0.32\linewidth}
        \centering
        \centerline{\includegraphics[width=3cm]{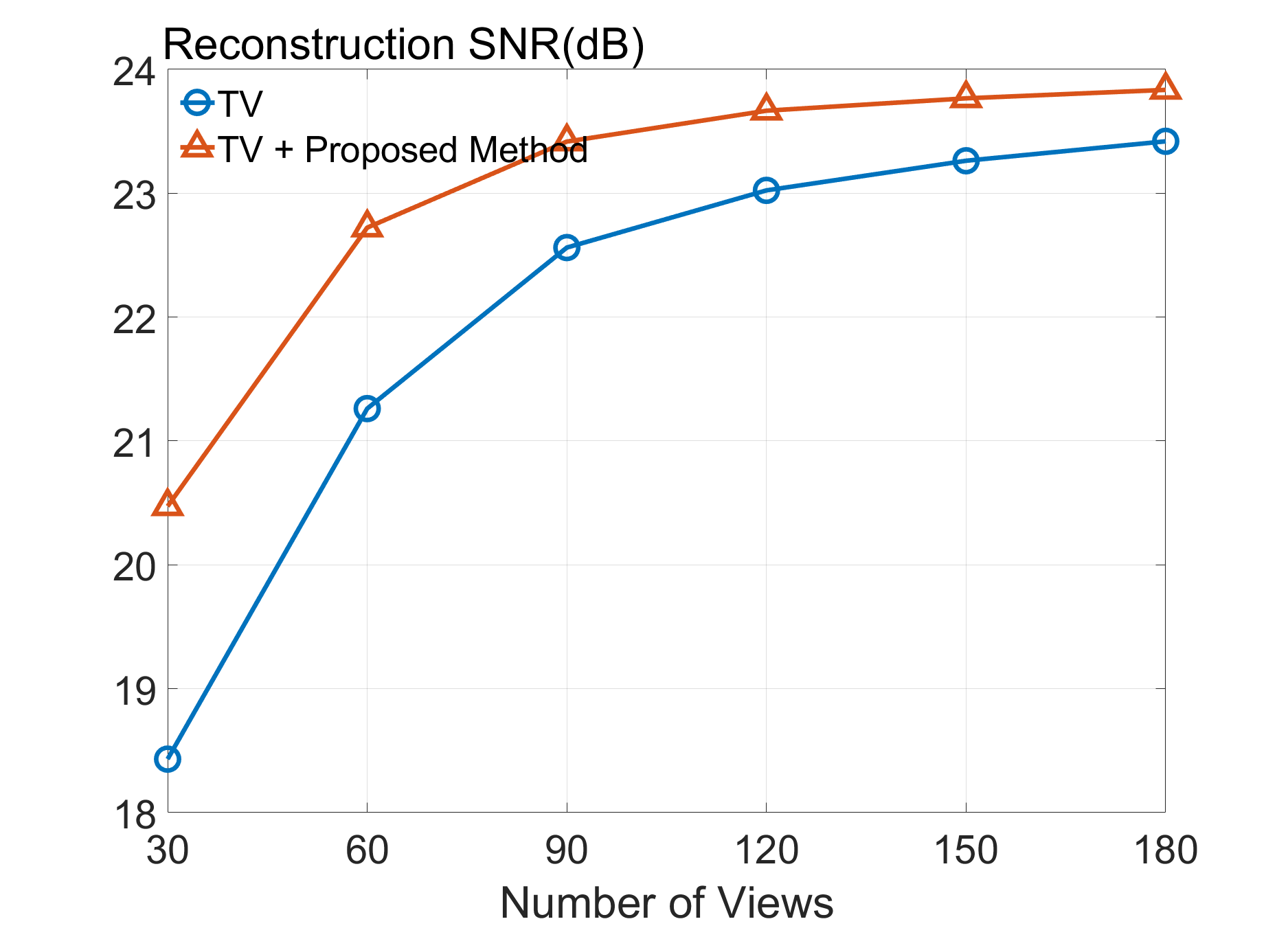}}
    \end{minipage}
    \begin{minipage}[b]{0.32\linewidth}
        \centering
        \centerline{\includegraphics[width=3cm]{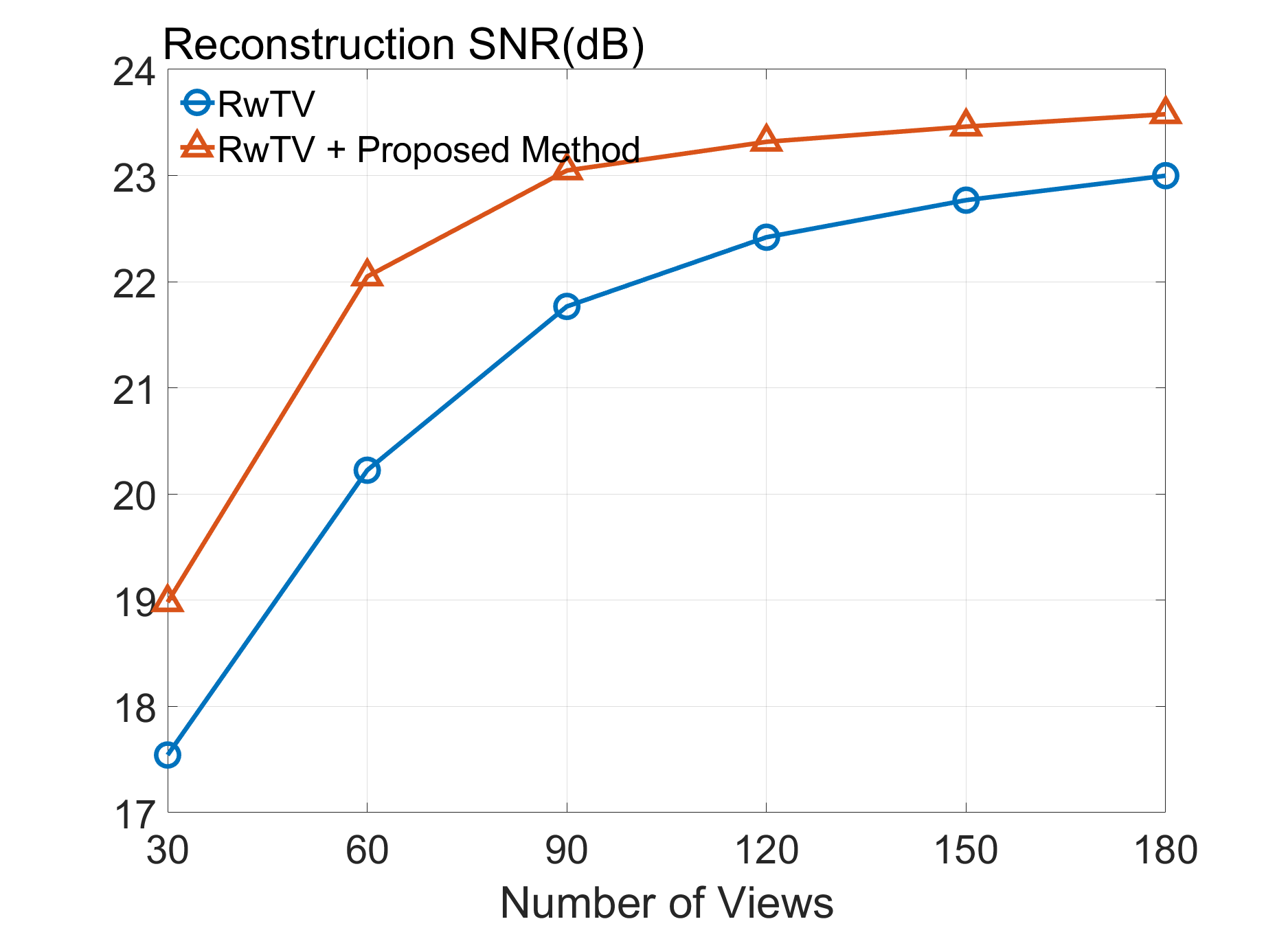}}
    \end{minipage}
    \begin{minipage}[b]{0.32\linewidth}
        \centering
        \centerline{\includegraphics[width=3cm]{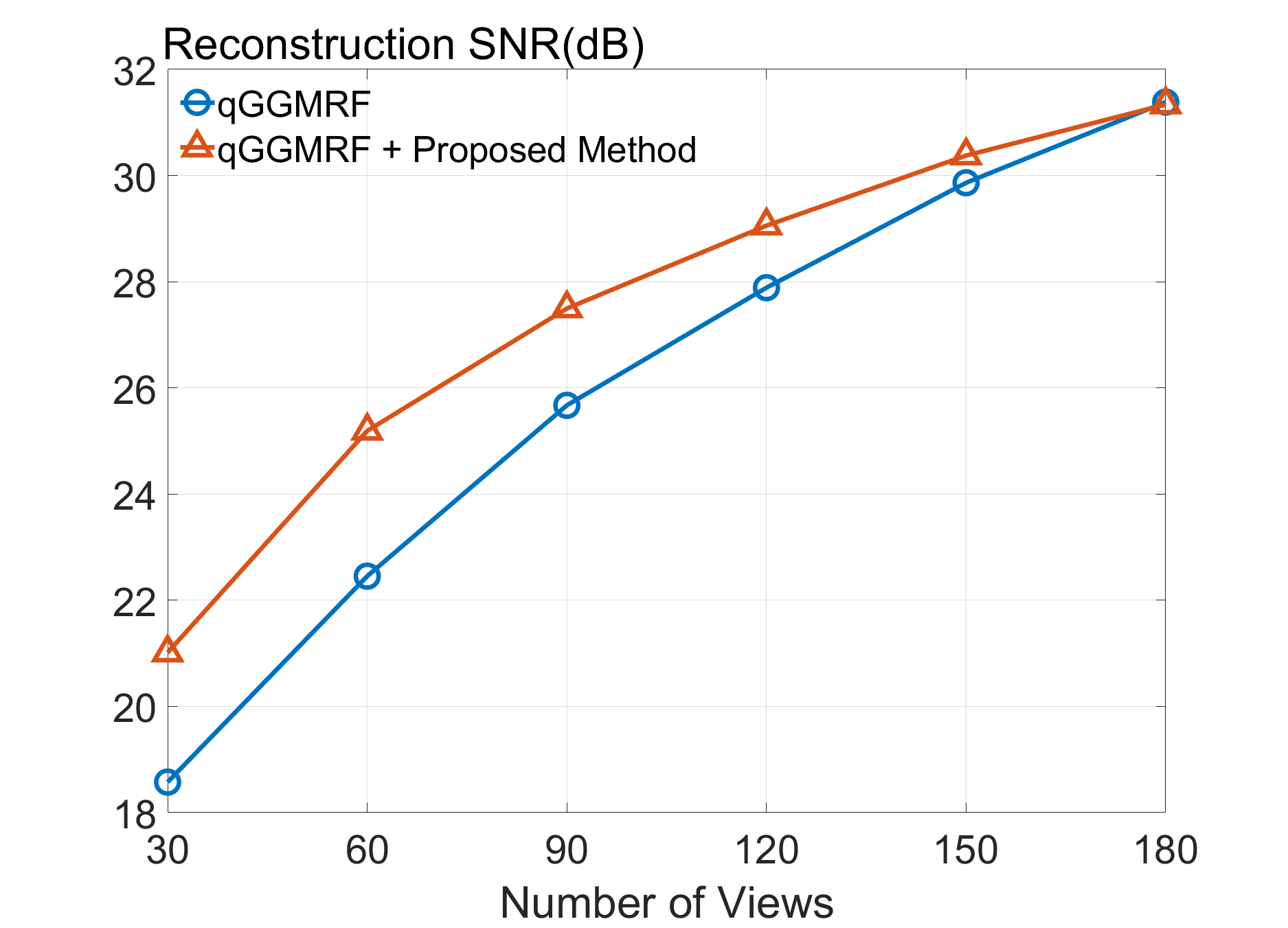}}
    \end{minipage}
    \caption{SNR(dB) comparison of TV, RwATV, and qGGMRF constraints with and without the proposed method under sparse-view conditions. The first row, the second row, and a third row correspond to the Shepp-Logan phantom, a brain CT image from HNSCC-3DCT-RT dataset, and the lung CT image from LoDoPaB-CT dataset respectively.}
    \label{snrsp}
\end{figure}

\begin{figure}[pos=htb]
    \begin{minipage}[b]{0.32\linewidth}
        \centering
        \centerline{\includegraphics[width=3cm]{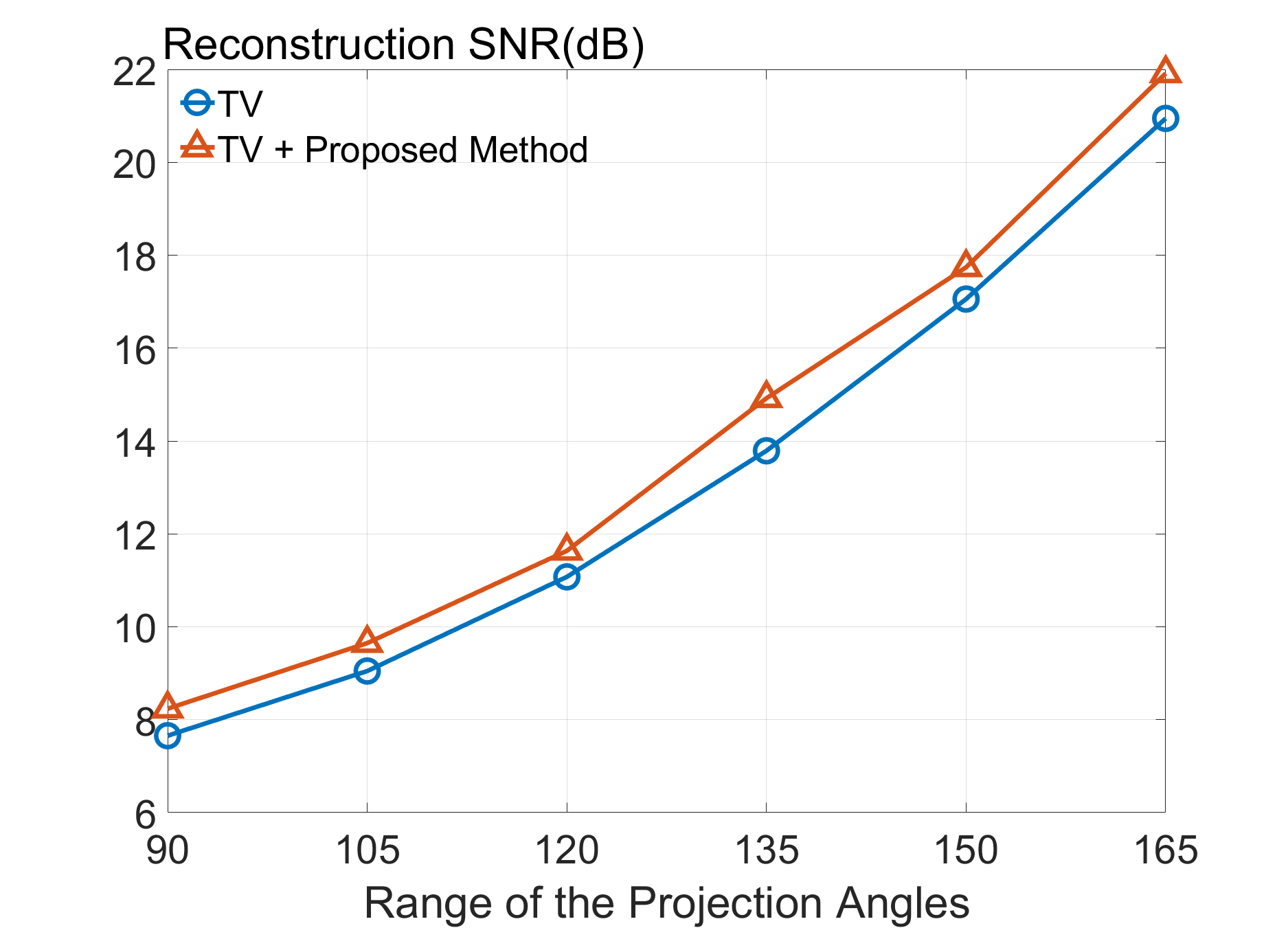}}
    \end{minipage}
    \begin{minipage}[b]{0.32\linewidth}
        \centering
        \centerline{\includegraphics[width=3cm]{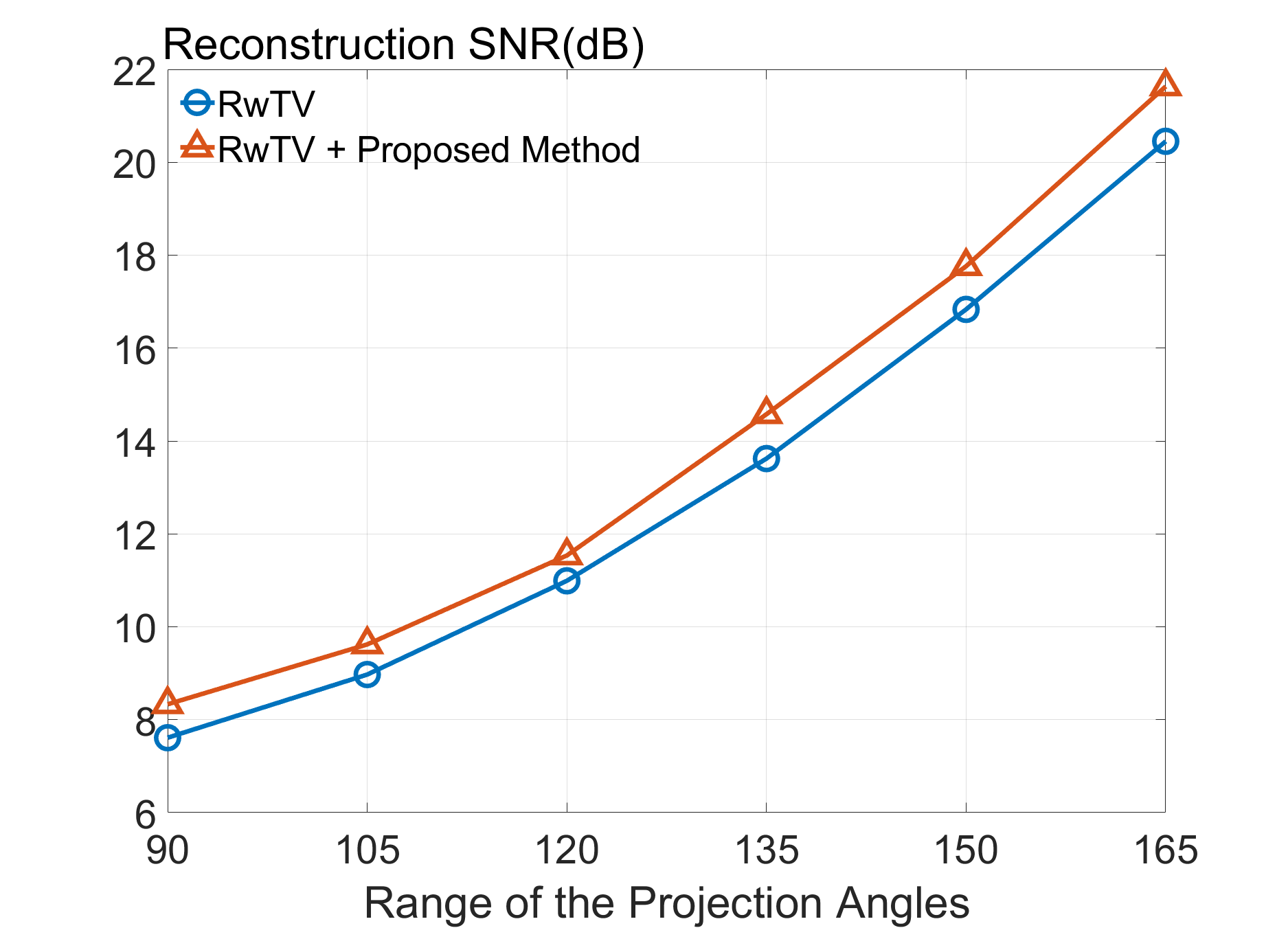}}
    \end{minipage}
    \begin{minipage}[b]{0.32\linewidth}
        \centering
        \centerline{\includegraphics[width=3cm]{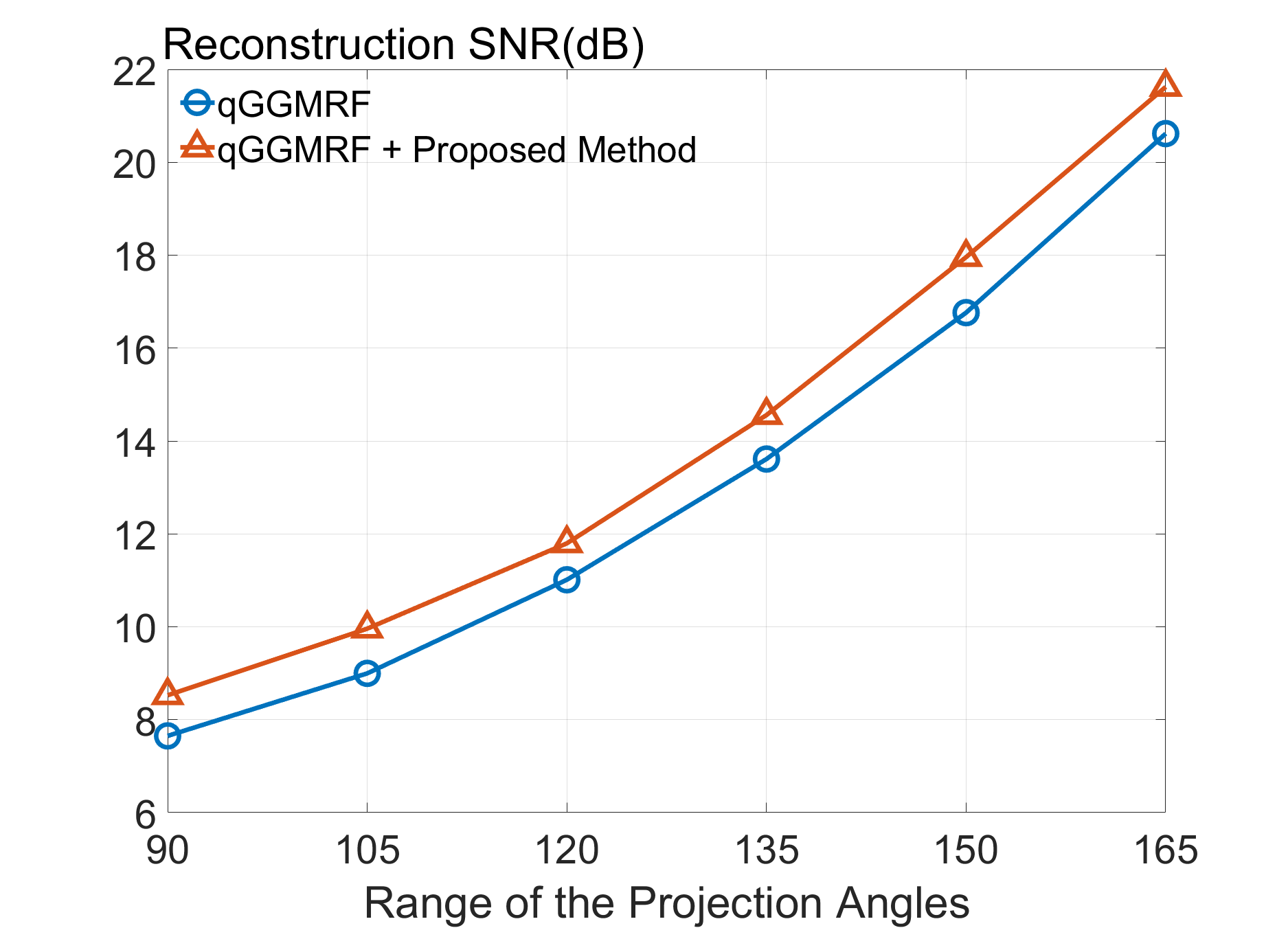}}
    \end{minipage}
    
    \begin{minipage}[b]{0.32\linewidth}
        \centering
        \centerline{\includegraphics[width=3cm]{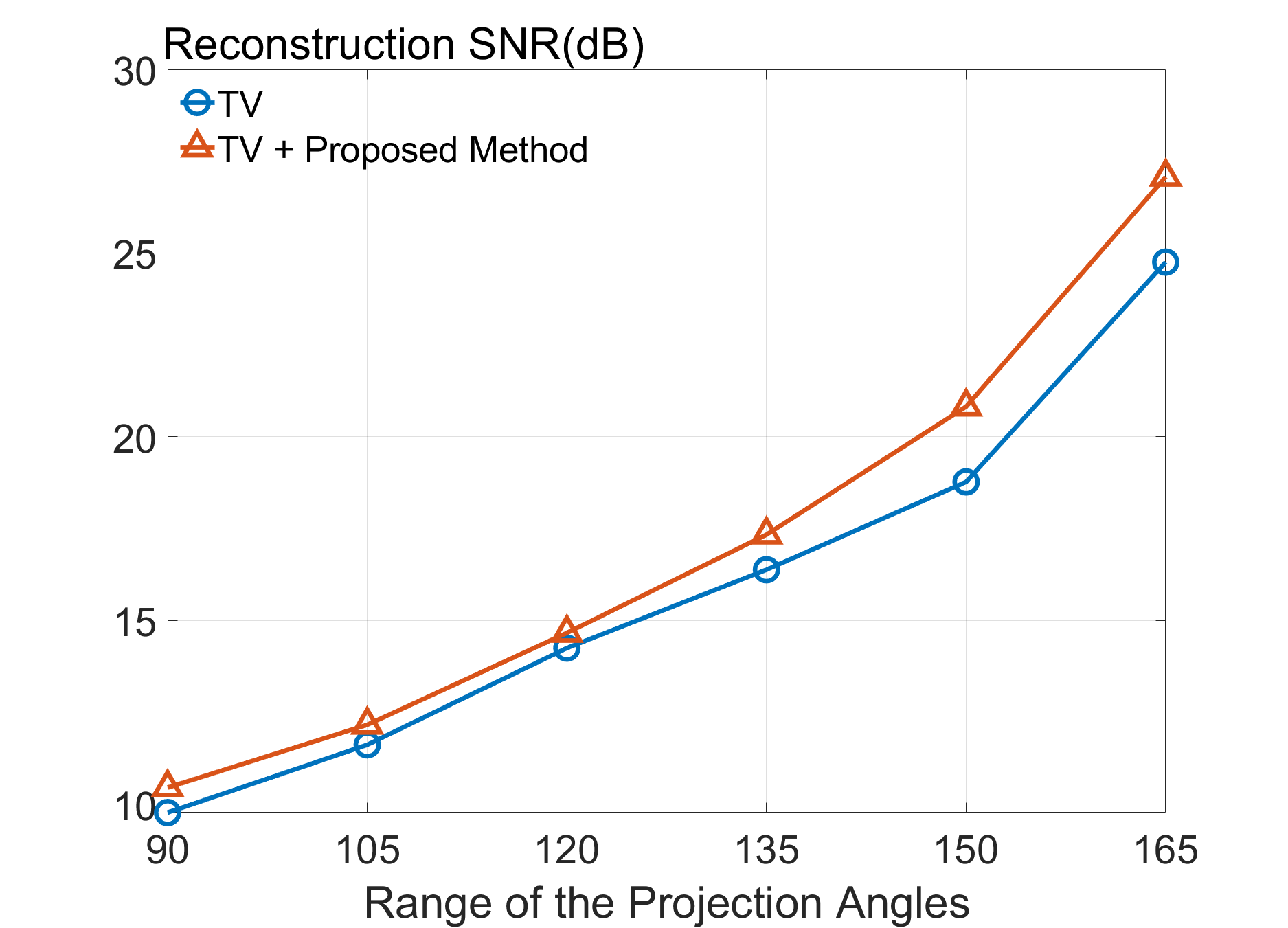}}
    \end{minipage}
    \begin{minipage}[b]{0.32\linewidth}
        \centering
        \centerline{\includegraphics[width=3cm]{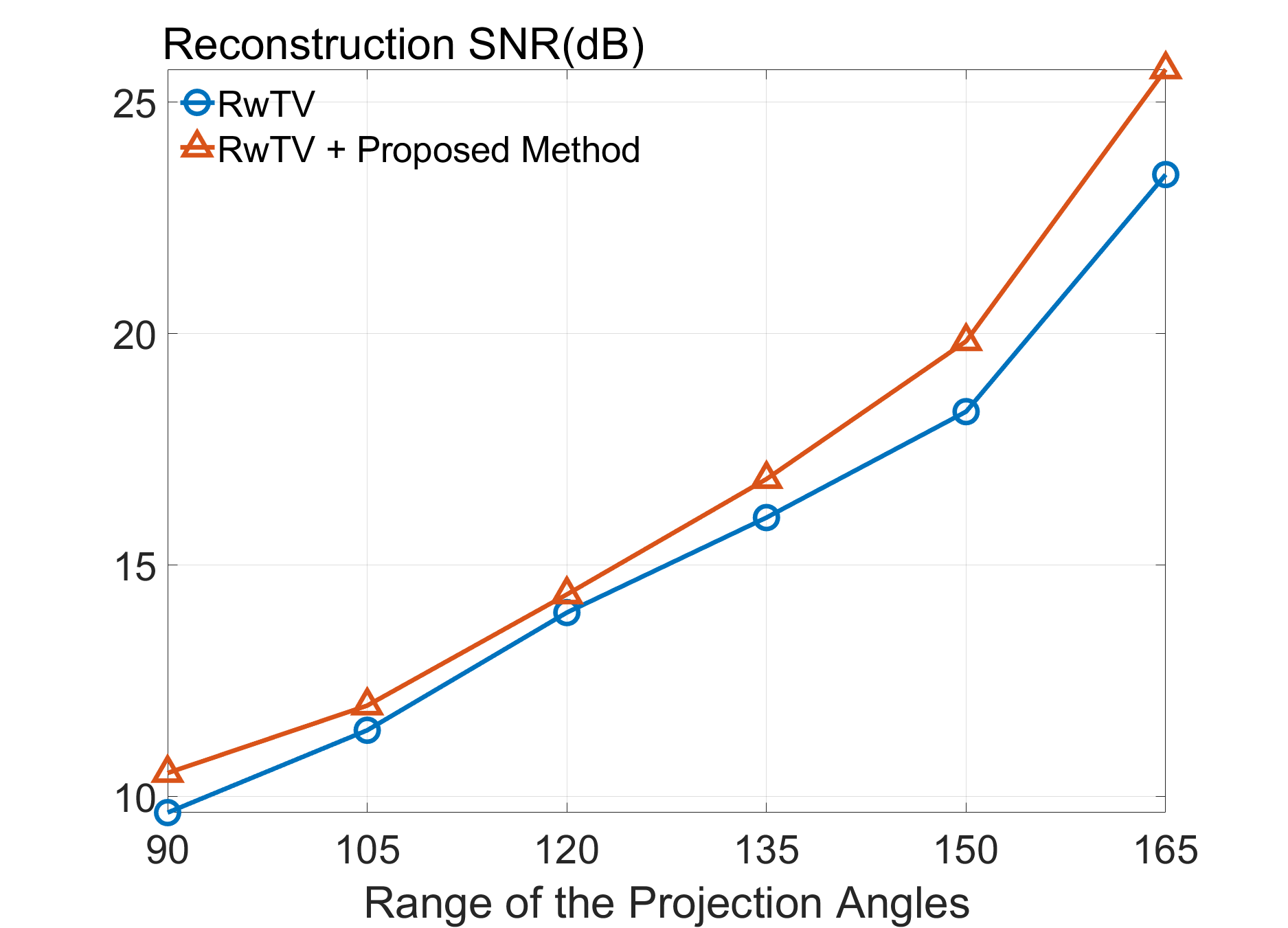}}
    \end{minipage}
    \begin{minipage}[b]{0.32\linewidth}
        \centering
        \centerline{\includegraphics[width=3cm]{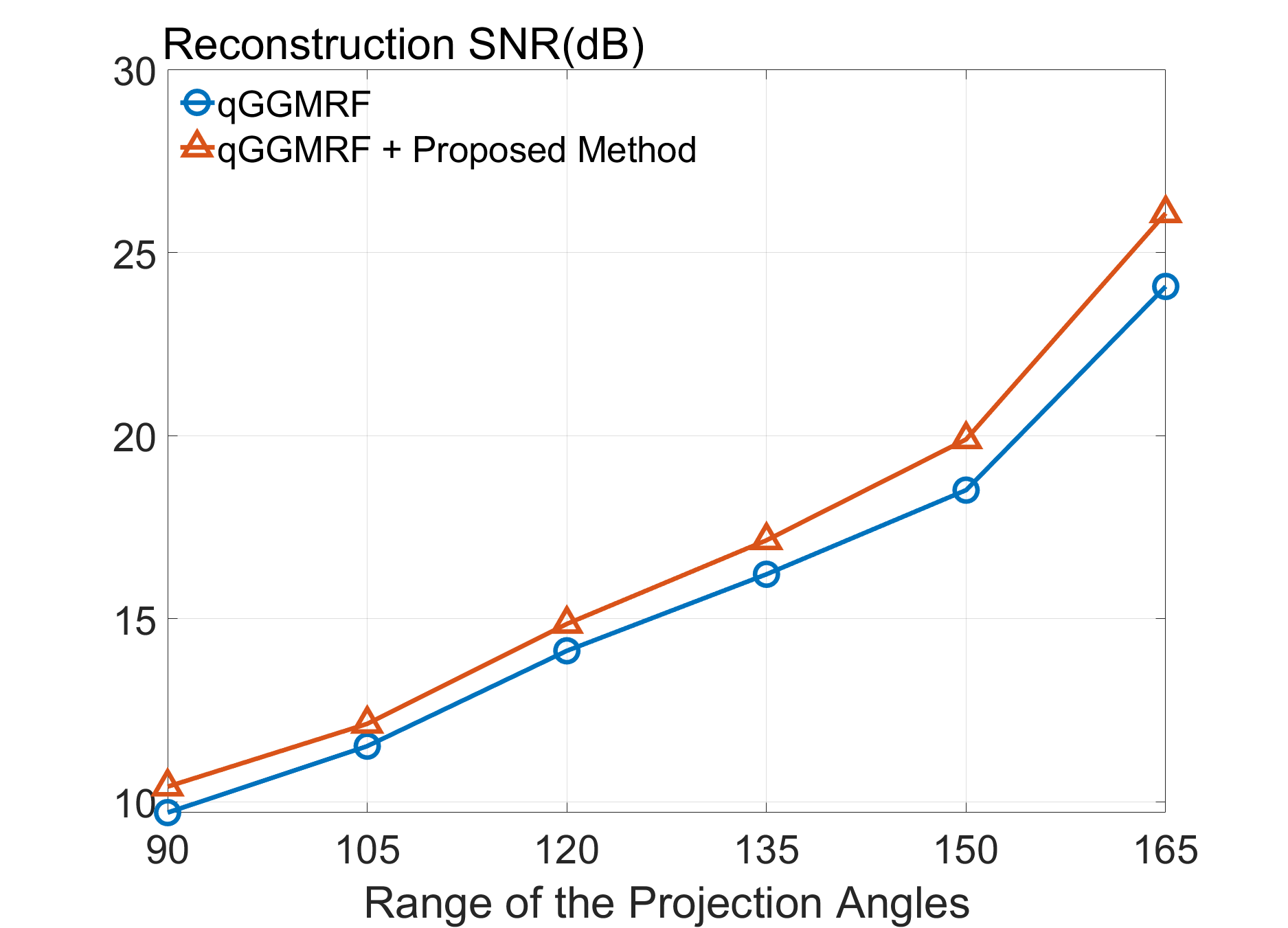}}
    \end{minipage}
    
    \begin{minipage}[b]{0.32\linewidth}
        \centering
        \centerline{\includegraphics[width=3cm]{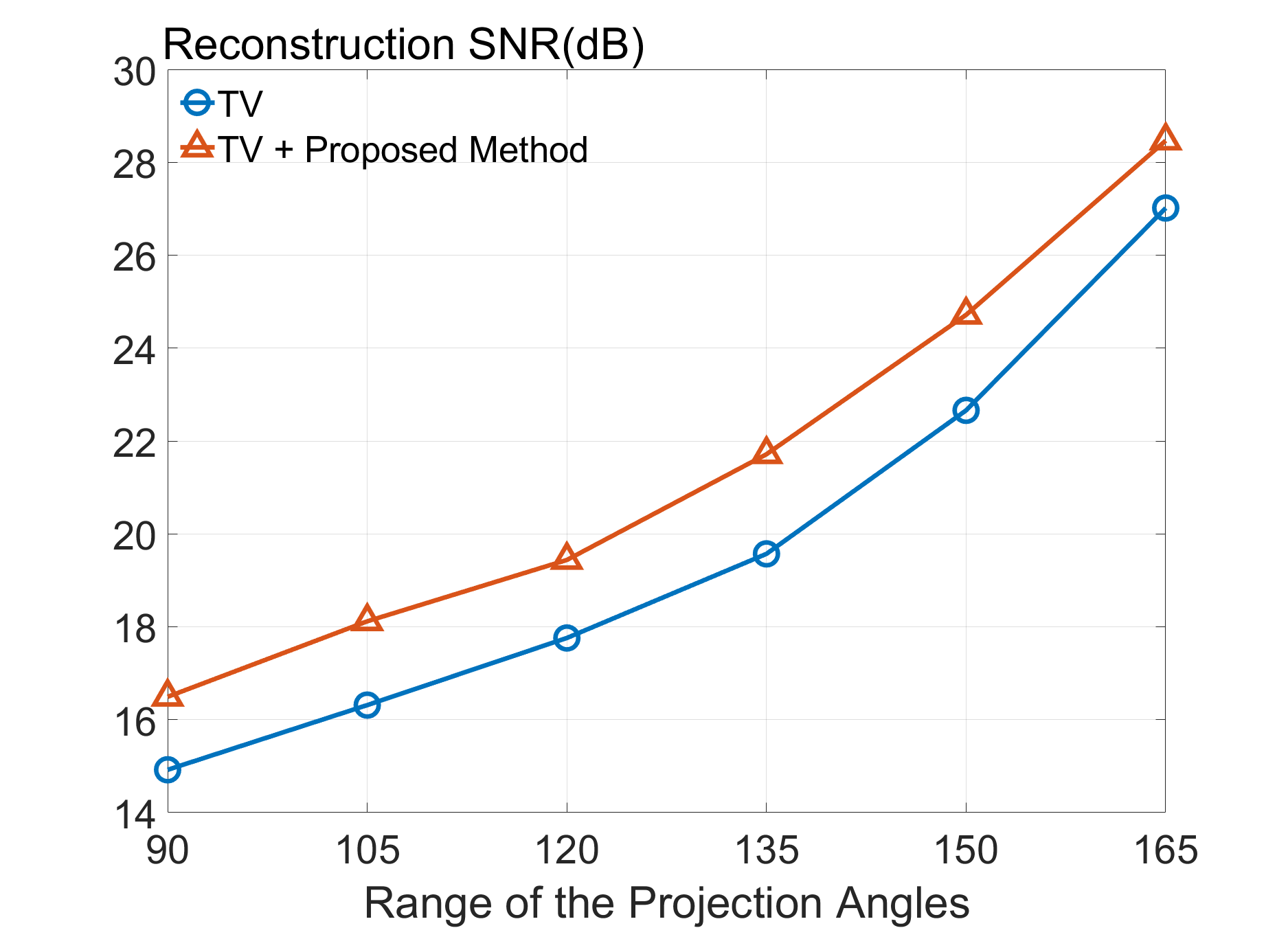}}
    \end{minipage}
    \begin{minipage}[b]{0.32\linewidth}
        \centering
        \centerline{\includegraphics[width=3cm]{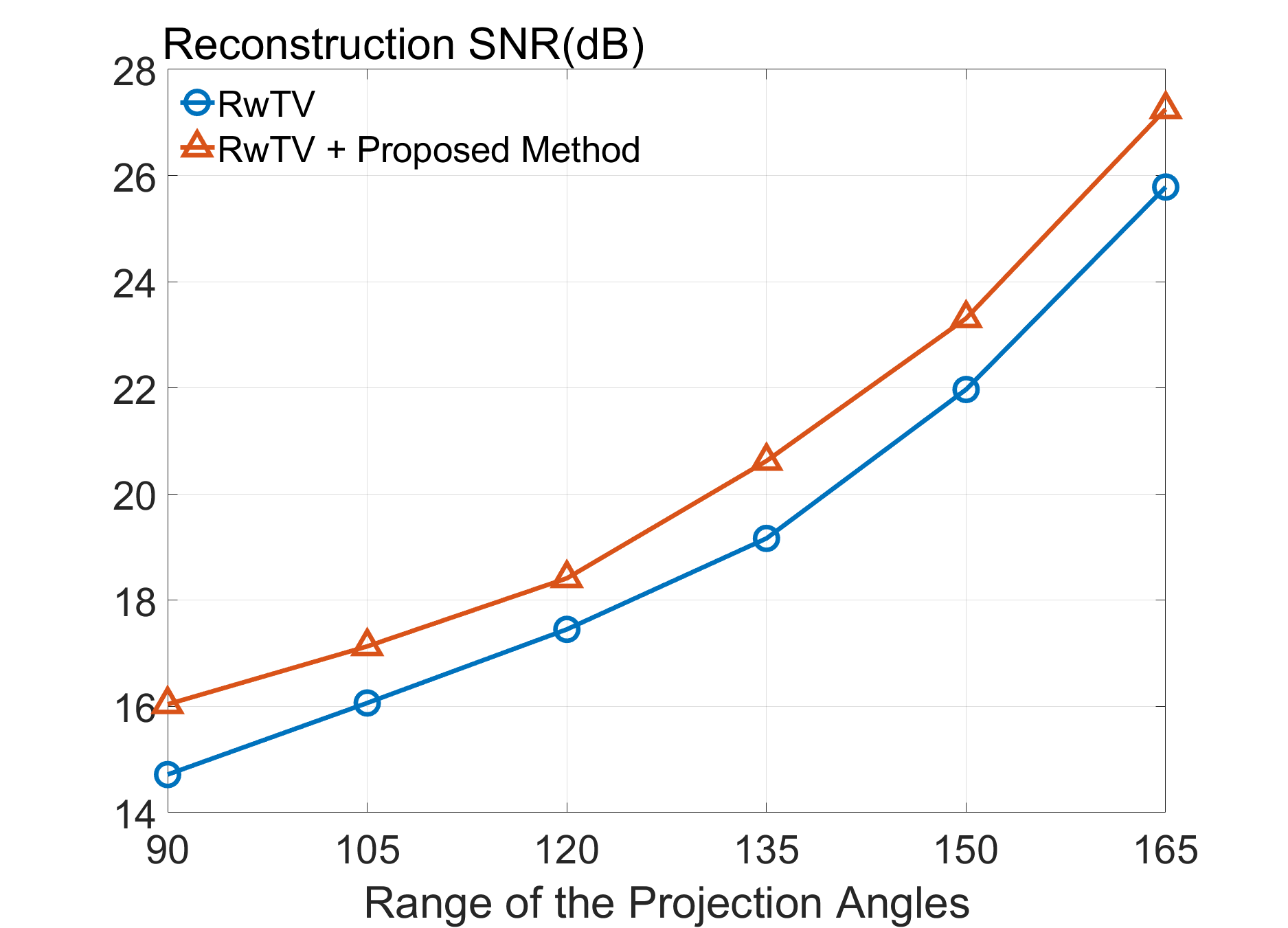}}
    \end{minipage}
    \begin{minipage}[b]{0.32\linewidth}
        \centering
        \centerline{\includegraphics[width=3cm]{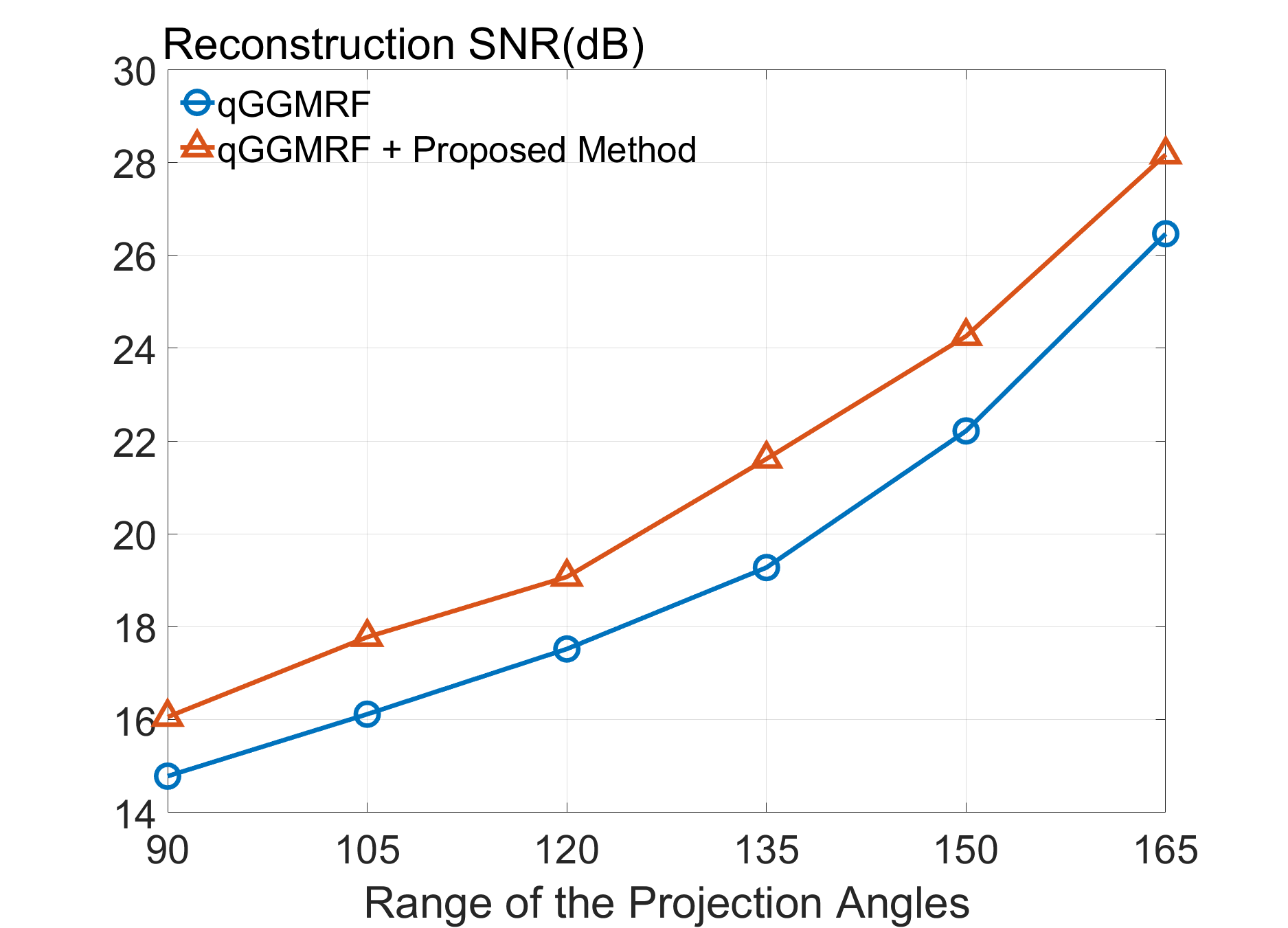}}
    \end{minipage}
    \caption{SNR(dB) comparison of TV, RwATV, and qGGMRF constraints with and without the proposed method under limited-angle conditions. The first row, the second row, and the third row correspond to the Shepp-Logan phantom, a brain CT image from HNSCC-3DCT-RT dataset, and a lung CT image from LoDoPaB-CT dataset respectively.}
    \label{snrla}
\end{figure}

To evaluate the improvement of the proposed method, the reconstruction accuracies in terms of signal-to-noise ratio (SNR) for TV, RwATV, and qGGMRF constraints with and without the proposed method are tested under sparse-view and limited-angle conditions. For sparse-view CT reconstruction, the projection angles uniformly distribute from $0^\circ$ to $180^\circ$, and the number of projections goes from $30$ to $180$. The performance of the reconstruction is shown in Fig.\ref{snrsp}, and some of the reconstruction results are shown in Fig.\ref{resultsp}. For limited-angle CT reconstruction, the angular range varies from $90^\circ$ to $165^\circ$ with one projection per degree. The performance of the reconstruction is shown in Fig.\ref{snrla}, and some of the reconstruction results are shown in Fig.\ref{resultla}. It is worth mentioning that for sparse-view CT reconstruction, the parameters $a$ and $b$ for the RwATV are set to $a = b = 1$ as the projections are isotropic.

\subsection{Noise in the Sinogram}\label{sec:noise}
\begin{figure}[pos=h]
    \begin{minipage}[b]{0.32\linewidth}
    \centering
    \centerline{\includegraphics[width=3cm]{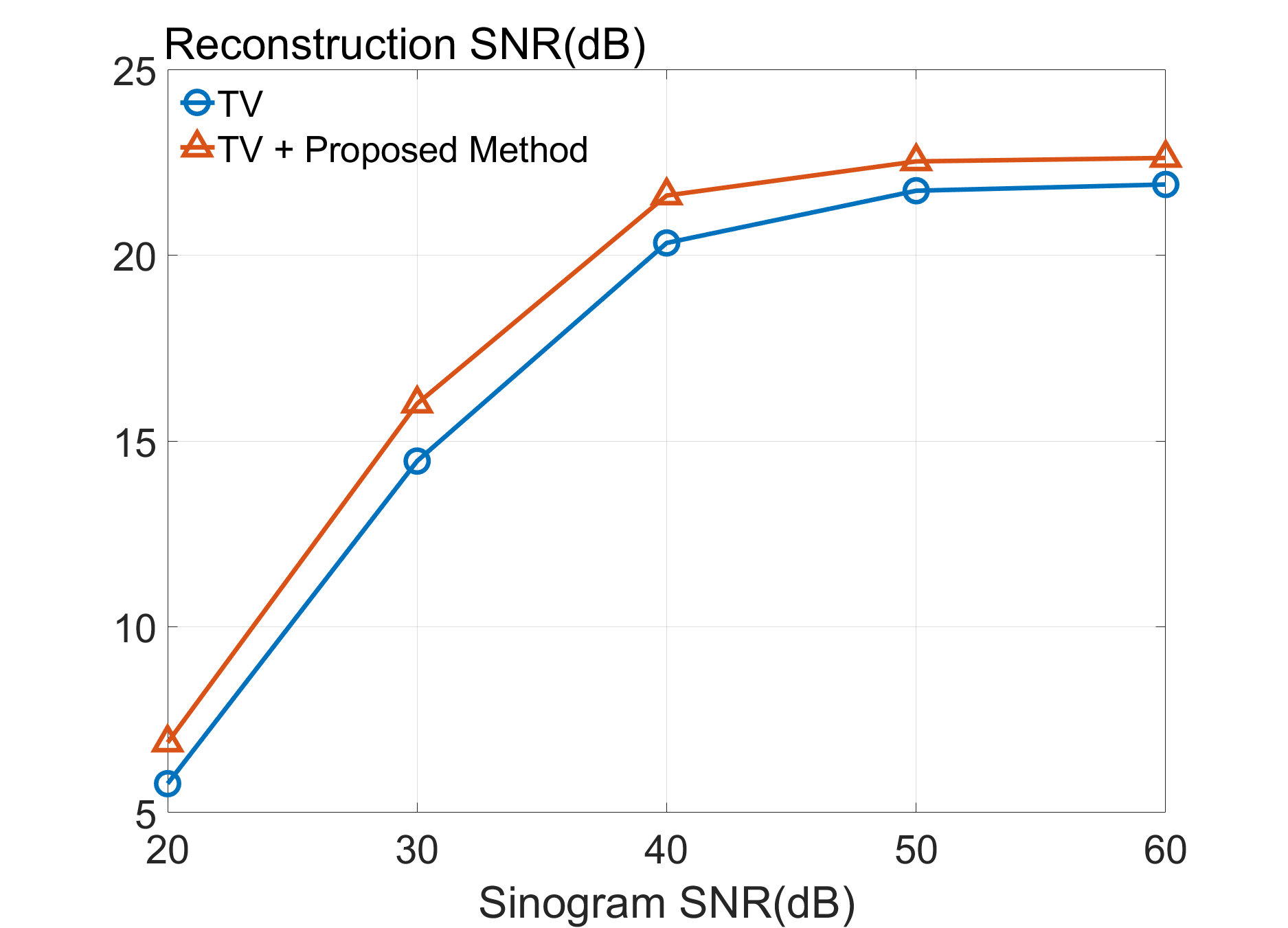}}
    \centerline{(a)}\medskip
\end{minipage}
\begin{minipage}[b]{0.32\linewidth}
    \centering
    \centerline{\includegraphics[width=3cm]{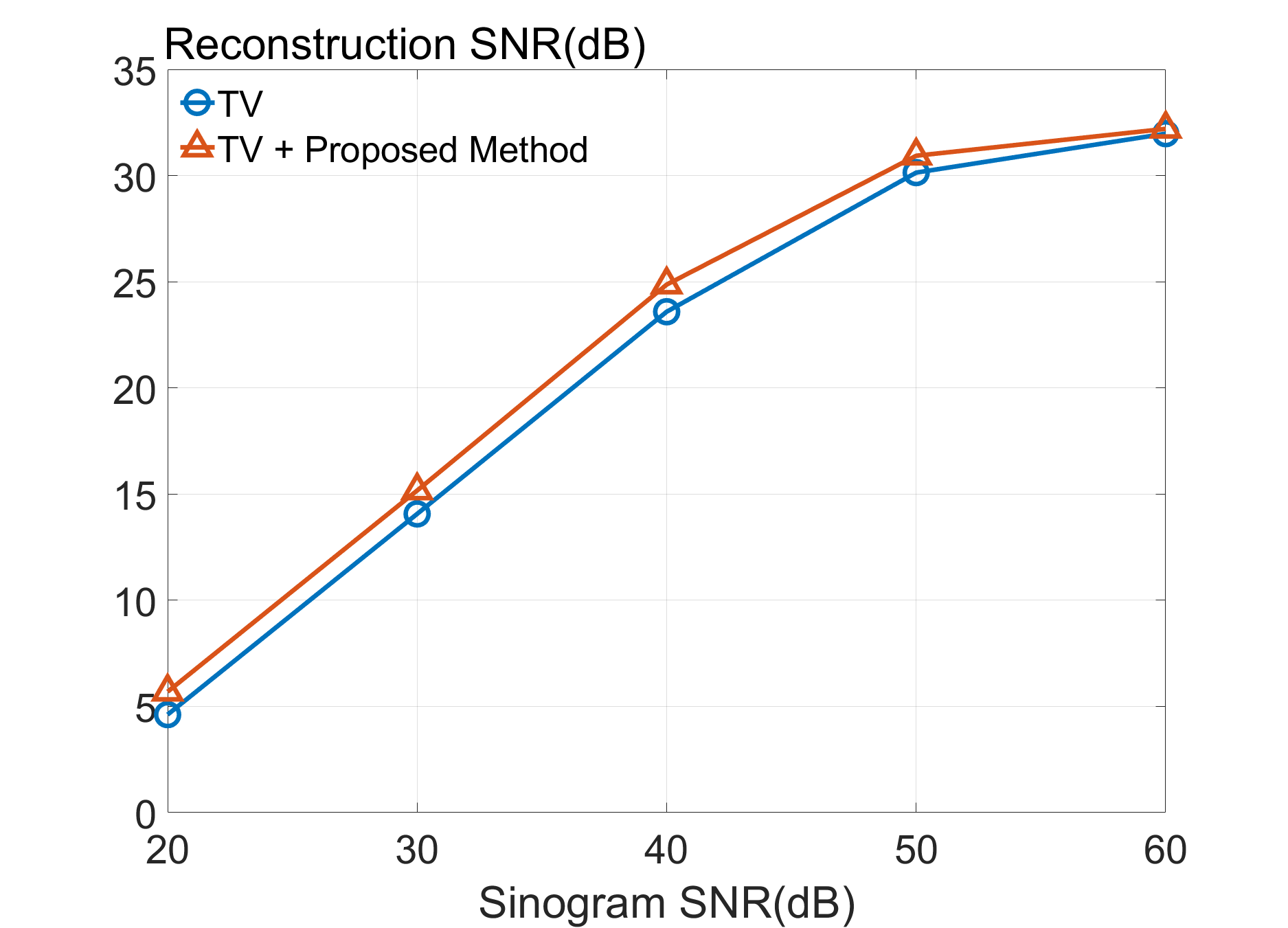}}
    \centerline{(b)}\medskip
\end{minipage}
\begin{minipage}[b]{0.32\linewidth}
    \centering
    \centerline{\includegraphics[width=3cm]{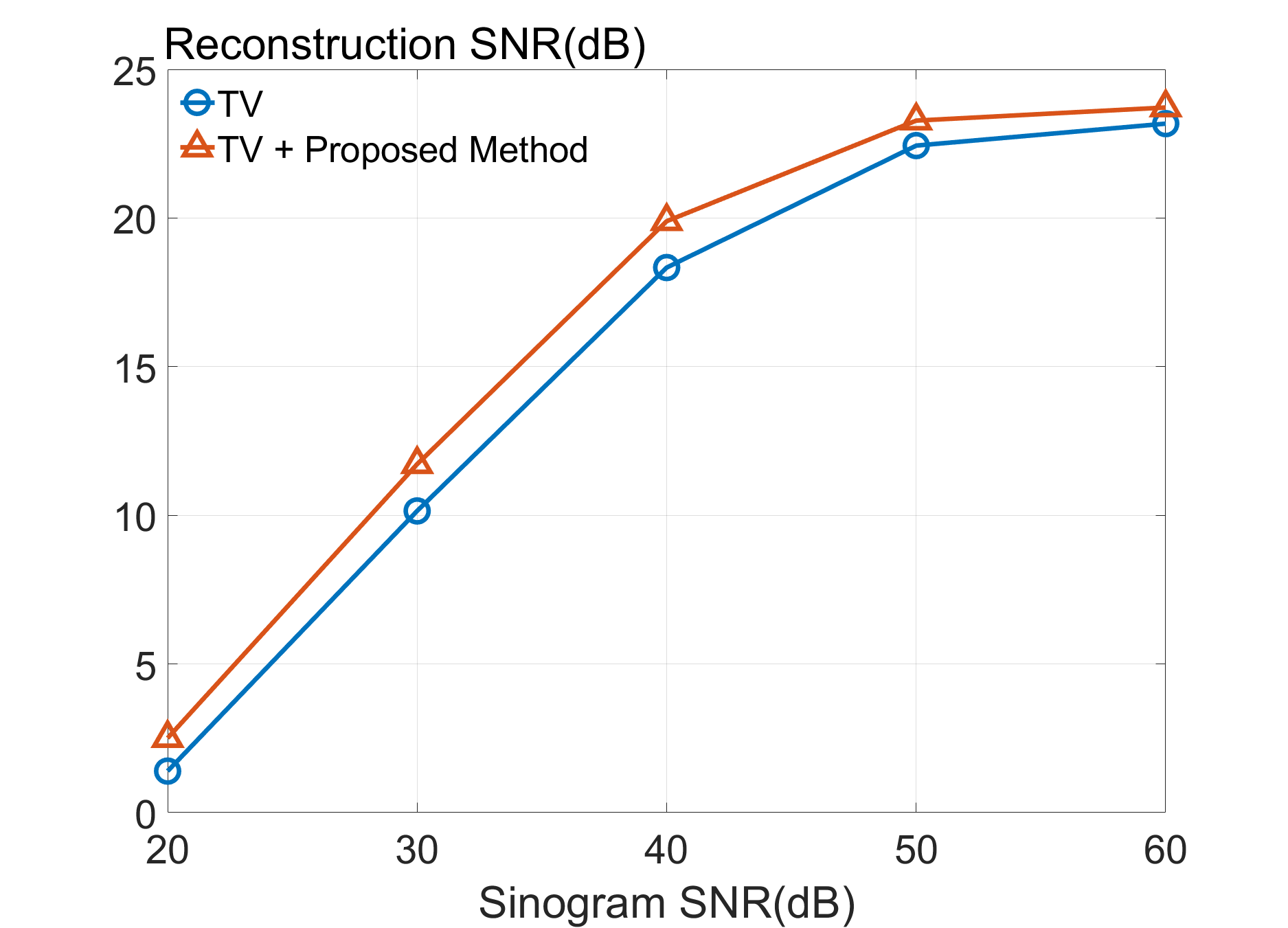}}
    \centerline{(c)}\medskip
\end{minipage}
\caption{The reconstruction SNR(dB) under low dose condition. (a), (b), and (c) correspond to the Shepp-Logan phantom, a brain CT image from HNSCC-3DCT-RT dataset, and a lung CT image from LoDoPaB-CT dataset respectively.}
\label{noise}
\end{figure}
In this section, we use the sinogram data polluted with Poisson distributed noise to test the proposed method's performance under low-dose conditions. The average number of X-ray photons received by the $i$th detector can be expressed as:
\begin{equation}
    E_i = I_0e^{[\mathcal{P}f]_i},
\end{equation}
where $I_0 > 0$ is the blank measurement ($[\mathcal{P}f]_i = 0$). It is worth mentioning that the sinogram data in our experiment is simulated by the Radon transform instead of acquired from a real instrument. Thus, $I_0$ here is a parameter for relative measurement. The experiment results on real images and phantoms are shown in Fig.\ref{noise}, where $I_0$ is set to $10^3, 10^4, 10^5, 10^6$, and $10^7$ corresponding to sinogram SNR $20$dB, $30$dB, $40$dB, $50$dB, and $60$dB.

\section{Discussion}
\label{sec:diss}
Recent research in this area focuses on increasingly complicated variants of TV (e.g. ATV, RwATV), or utilizing more neighbors (e.g. 8 neighbors and 16 neighbors) with delicate weights at different distances and directions. However, these approaches see diminishing returns as they all utilize the piece-wise constant prior.

In this study, we point out that the approximated intensity of each composition in CT images can also be used as a prior. With this prior, we propose a method to further improve CT reconstruction under non-ideal conditions. Furthermore, the proposed method can collaborate with most existing constraints as it only utilizes the global information of CT images. The differences between the proposed method and the DART related algorithms are the non-discrete imaged object and unknown composition numbers. To overcome these challenges, the number of compositions in the proposed method is set to increase along with the number of iterations. Its effectiveness is shown in the experiment in Section \ref{sec:3.1}. Fig.\ref{reconprocess}a illustrates that streak artifacts and blurred effects can be greatly reduced by gray level segmentation in the early iteration. Fig.\ref{reconprocess}b and Fig.\ref{reconprocess}c show that the increase of composition number ($n$) does help generate detailed non-discrete CT images, and will not downgrade the reconstruction accuracy even if $n$ is larger than optimal. 

As mentioned before, a high-quality reconstruction cannot be obtained by simply minimizing the data inconsistency since the back projection related algorithms cannot affect frequency pixels in the unmeasured area. Furthermore, as shown in Fig.\ref{reconprocess2}b, the magnitude of the update made by the IR algorithm decreases drastically. This implies that increasing the number of iterations cannot effectively increase the accuracy. This problem is solved by the proposed method. The effect of the proposed method is not to increase SNR directly, but to steer IR algorithms to a better result. The update made by the proposed method may not be fully correct, but its incorrect part can be rectified by IR algorithms in the later iteration, and its correct part always corresponds to the frequency pixels in the unmeasured areas, where IR algorithm cannot modify. This is in line with Fig.\ref{reconprocess2}: for every $50$ iteration, the SNR of the proposed method will slightly decrease and the magnitude of IR update will drastically increase, resulting in a $2$dB improvement.

As a result, the proposed method can further improve the reconstruction accuracy by about $2$dB under multiple conditions (sparse-view, limited-angle, and low-dose) for multiple imaged objects (phantom, brain, and lung). The corresponding performance plots are available in Fig.\ref{snrsp}, Fig.\ref{snrla}, and Fig.\ref{noise}, some of the reconstruction results are shown in Fig.\ref{resultsp} and Fig.\ref{resultla}. In order to minimize the error caused by discretization and the model ($\boldsymbol H$), only the parallel beam geometry with pixel basis~\cite{shu2020gram,shu2022exact} is tested. However, it is evident that the proposed method can be used in any basis and geometries or even other medical imaging areas such as MRI.

\section{Conclusions}
\label{sec:conclusion}
 In this paper, we propose a method utilizing the prior of imaged objects' composition to further improve the accuracy of CT reconstruction under non-ideal conditions. Such methods can collaborate with most local constraints as these constraints focus on the local relationship between pixels and their neighbors. In our experiments, the proposed method is tested with TV, RwATV, and qGGMRF, all state-of-the-art constraints for sparse-view, limited-angle, and low-dose CT reconstruction. The experiments show significant and stable improvement under multiple circumstances. With these results, we conclude that the proposed method is effective for further improving the performance of limited-angle CT reconstruction. Also, the proposed method still has great potential for further improvement. Although Otsu's method is used in this paper, it is evident that better performance can be obtained if more delicate segmentation algorithms are used.

\section*{Declaration of Competing Interest}
The authors declare that they have no known competing financial interests or personal relationships that could have appeared to influence the work reported in this paper.

\bibliographystyle{cas-model2-names}
\bibliography{strings,refs}

\end{document}